# Alternating Deep Low-Rank Approach for Exponential Function Reconstruction and Its Biomedical Magnetic Resonance Applications

Yihui Huang, Zi Wang, Xinlin Zhang, Jian Cao, Zhangren Tu, Meijin Lin, Di Guo, Xiaobo Qu*

*Abstract*—Undersampling can accelerate the signal acquisition but at the cost of bringing in artifacts. Removing these artifacts is a fundamental problem in signal processing and this task is also called signal reconstruction. Through modeling signals as the superimposed exponential functions, deep learning has achieved fast and high-fidelity signal reconstruction by training a mapping from the undersampled exponentials to the fully sampled ones. However, the mismatch, such as the sampling rate of undersampling, the organ and the contrast of imaging, between the training and target data will heavily compromise the reconstruction. To address this issue, we propose Alternating Deep Low-Rank (ADLR), which combines deep learning solvers and classic optimization solvers. Experiments on the reconstruction of synthetic and realistic biomedical magnetic resonance signals demonstrate that ADLR can effectively mitigate the mismatch issue and achieve lower reconstruction errors than state-of-the-art methods.

## I. Introduction

### A. Applications of Exponential Functions

In many signals processing, signals can be expressed by the exponential function in many scenarios, such as radar [1], sonar [2], analog-to-digital converter [3], single event-related potentials in the brain [4] and signal from magnetic resonance (MR) [5-7]. As one of the most valuable modern biomedical tools, the MR technology can analyze protein characteristics [5, 7-12], anatomical structure [13-16] and the spatial distribution of cellular metabolites [17, 18]. Bloch equation describes the generation mechanism of MR signals, which can be expreesed by exponential functions [5, 6]. Thus, utilizing the characteristic of exponentials is important in the applications of biomedical MR [5, 8-10, 14, 19, 20].

In general, the $n^{th}$ sampled data point of the MR signals $\mathbf{x} \in \mathbb{C}^N$, with a sampling interval $\Delta t$, can be approximately expressed by the superposition of $G$ exponential components[7]:

$$\mathbf{x}(n\Delta t) = \sum_{g=1}^{G} (A_g e^{i\phi_g}) e^{-\frac{n\Delta t}{\tau_g}} e^{i2\pi f_g n\Delta t}, \qquad (1)$$

where the indices of sampled signals and exponential components are respectively denoted by $n = 0, 1, 2, ..., N-1$ and $g = 1, 2, 3, ..., G$. In addition to the amplitude $A_g$ and the damping factor $\tau_g$ being the positive numbers, the range of normalized frequency $f_g$ is from 0 to 1 and the initial phase $\phi_g$ is from 0 to $2\pi$.

### B. Challenges in Signal Reconstruction

Undersampling can accelerate the MR signal acquisition but at the cost of leading to artifacts, which is an important issue in biomedical applications [5], such as fast nuclear MR (NMR) spectroscopy [8-10, 19-22] and fast MR imaging (MRI) [13-16]. The undersampled signal $\mathbf{y} \in \mathbb{C}^M$ with $M$ ($< N$) sampling points is modeled as $\mathbf{y} = \mathcal{U}\mathbf{x} + \mathbf{n}$, where $\mathcal{U}$ is an undersampling operator with a sampling rate $M$ over $N$, and $\mathbf{n} \in \mathbb{C}^M$ denotes Gaussian noise. Due to $M < N$, the reconstruction of the original $\mathbf{x}$ is relatively hard due to the underdetermined system of equations. A powerful way is to introduce some priors, such as sparsity [21, 22] or low rank [9, 23], to constrain the solution.

The sparsity prior is introduced to minimize the number of nonzero values in the MR signals. With slight or no damping, the signal of narrow spectral peaks can be recovered faithfully by compressed sensing [21, 22]. However, the sparsity in the frequency domain is not satisfied well if spectral peaks are broad when the damping decay is fast, which raises challenges in practical applications [9, 23, 24].

To break the limitation of broad spectral peaks in the reconstruction of MR signals, the number of exponential functions can be minimized according to the low-rank prior [9, 10, 17, 23, 25]. The time-domain signals can construct one Hankel matrix, whose rank is the same as the number of exponential components [11]. Given the number of exponential components is less than the matrix size, the matrix would be low rank and this property has been used to faithfully reconstruct undersampled MR signals [8-10]. However, the constraint of low rank is usually realized by minimizing the sum of matrix singular values, resulting in slow computation because singular value decomposition (SVD) [26] costs much

This work was partially supported by the National Natural Science Foundation of China (62331021, 62122064 and 62371410), the Natural Science Foundation of Fujian Province of China (2023J02005 and 2021J011184), Industry-University Cooperation Projects of the Ministry of Education of China (231107173160805), National Key R&D Program of China (2023YFF0714200), the President Fund of Xiamen University (20720220063), Nanqiang Outstanding Talent Program of Xiamen University.

Yihui Huang, Zi Wang, Jian Cao, Zhangren Tu and Xiaobo Qu are with the Department of Electronic Science, Biomedical Intelligent Cloud Research and Development Center, Fujian Provincial Key Laboratory of Plasma and Magnetic Resonance, Xiamen University, Xiamen, China (*Corresponding author, email: quxiaobo@xmu.edu.cn).

Xinlin Zhang is with the College of Physics and Information Engineering, Fuzhou University, Fuzhou, China

Meijin Lin is with the Department of Applied Marine Physics and Engineering, College of Ocean and Earth Sciences, Xiamen University

Di Guo is with the School of Computer and Information Engineering, Xiamen University of Technology, Xiamen, China.

time. Thus, LRHMF [8] utilizes the SVD-free algorithm [27] to reduce the reconstruction time in each iteration. Nevertheless, too many iterations in LRHMF still lead to a relatively long reconstruction time [19].

### C. Deep Learning Approaches

Deep learning [28-30] is a prominent technology that offers the appealing advantage of powerful learning ability and high computation speed. DLNMR [19] and FID-Net[31] show the potential of deep learning to reconstruct MR signals and is applied to fast NMR spectroscopy. According to Eq.(1), neural networks can be trained by MR signals composed purely of exponentials [5]. Moreover, the introduction of the low-rank prior constrains the network through the unrolling design of LRHMF [20], which is called deep Hankel matrix factorization (DHMF). It decreases the reconstruction error for weak spectral peaks and improves the interpretability of deep learning.

Notably, deep learning is trained on the dataset of the MR signals that share some same properties, such as the sampling rate, the number of peaks and the length of signals. For deep learning, the nature of training aims to learn how to get the best performance of the training dataset. When target signals have some properties that are different from the training set, called the mismatch between target and training signals, the quality of the reconstructed signal deteriorates [32]. Taking the mismatched sampling rate as an example (Fig. 1), DHMF [20] introduces severe distortions in the reconstruction of a target spectrum with a sampling rate of 50% when it is trained with a sampling rate of 25%. As a comparison, the traditional optimization solver, LRHMF, provides high-quality spectra (Fig. 1(b)(e)). This result implies that the data-based low-rank constraint cannot provide robust reconstruction while the traditional optimization solver still has great merit.

### D. Brief Introduction to the Proposed Method

To overcome the non-robust of deep learning that depends on the properties of training sets, one intuitive idea is to introduce data-free optimization solvers into deep learning network structures, so that the reconstructed signal is constrained well by an optimization solver. Therefore, low-rank prior is introduced and the merit of optimization solvers is inherited.

In our paper, an alternating deep low-rank approach (ADLR) is proposed, which consists of the deep learning solver and the optimization solver of low-rank prior, and reconstructs the MR signal faithfully even with the mismatching properties. This method is tested on the synthetic exponentials, realistic NMR spectroscopy data and MRI data in this work.

In brief, the paper's main contribution includes:

(1) To improve the robustness of the reconstruction of biomedical signals, an alternating structure of deep learning solvers and optimization solvers based on the low-rank property is proposed.

(2) Without retraining, the proposed method achieves a lower reconstruction error of biological nuclear magnetic resonance spectroscopy under different sampling rates or medical magnetic resonance imaging of different organs and image contrasts than the state-of-the-art methods.

## II. BACKGROUND

### A. Notations

For easy reading, some notations throughout the paper are defined: the scale, vector, matrix and tensor are represented by $x$, $\mathbf{x}$, $\mathbf{X}$ and $\mathcal{X}$. $\mathbb{C}$ and $\Omega$ are the sets, and $\mathbf{I}$ denotes an identity matrix. The superscript $H$ means the conjugate transpose of the matrix. $\mathbf{X}_{ij}$ means the number in the $i^{\text{th}}$ row and $j^{\text{th}}$ column of matrix $\mathbf{X}$. $\|\cdot\|_2$ denotes the $\ell_2$ norm of vectors, the $\|\cdot\|_*$ and $\|\cdot\|_F$ means the nuclear norm and Frobenius norm of the matrix.

Operators are represented by calligraphic letters. $\mathcal{U}$ stands for the undersampling operator that samples $M$ points from $N$ points ($M<N$). $\mathcal{U}^*$ denotes the corresponding zero-filling operator that fills zeros into the unsampled ($N$-$M$) data points.

For reconstruction of exponential function and NMR spectra, $\mathcal{H}$ denotes the Hankel operation [9, 11, 20] that transforms a vector $\mathbf{x}$ of length $N_1+N_2-1$ to a matrix $\mathbf{X}$ of size $N_1 \times N_2$:

$$\mathcal{H}: \mathbf{x} \in \mathbb{C}^{N_1+N_2-1} \to \mathbf{X} \in \mathbb{C}^{N_1 \times N_2} \text{ and } \mathbf{X}_{ij} = \mathbf{x}_{i+j-1}. \quad (2)$$

The operator $\mathcal{H}^*$ maps a matrix $\mathbf{X}$ to a vector $\mathbf{x}$ with the relationship $\mathcal{H}^*(\mathcal{H}\mathbf{x}) = \mathbf{x}$ as follows:

$$\mathcal{H}^*: \mathbf{X} \in \mathbb{C}^{N_1 \times N_2} \to \mathbf{x} \in \mathbb{C}^{N_1+N_2-1}, \mathbf{x}_{i+j-1} = \frac{\sum_{i+j-1} \mathbf{X}_{ij}}{\sum_{i+j-1}(\mathcal{H}\mathbf{o})_{ij}}, \quad (3)$$

where $\mathbf{o} \in \mathbb{C}^{N_1+N_2-1}$ is a vector of ones and $\sum_{i+j-1}(\mathcal{H}\mathbf{o})_{ij}$ is the number of entries in the anti-diagonal of matrix $\mathcal{H}\mathbf{o}$.

Similarly, for MRI reconstruction, $\mathcal{H}_{VC}$ denotes the Hankel operator with virtual coils [14, 33] that transforms a matrix $\mathbf{X}$ to another matrix $\mathbf{Y}$.

$$\mathcal{H}_{VC}: \mathbf{X} \in \mathbb{C}^{(N_1+N_2-1) \times N_3} \to \mathbf{Y} \in \mathbb{C}^{N_1 \times 2N_2 N_3},$$

$$\mathcal{H}_{VC}\mathbf{X} = [\mathcal{H}\mathbf{X}_{:,1}, \cdots, \mathcal{H}\mathbf{X}_{:,N_3}, \mathcal{H}(\mathbf{X}_{:,1})^\dagger, \cdots, \mathcal{H}(\mathbf{X}_{:,N_3})^\dagger], \quad (4)$$

where $\mathcal{H}(\mathbf{X}_{:,j})^\dagger$ with the superscript † denotes the flipping and conjugate of the vector $\mathbf{X}_{:,j}$ in the $j^{\text{th}}$ column of matrix $\mathbf{X}$ along the center. The corresponding operator $\mathcal{H}_{VC}^*$ is defined:

$$\mathcal{H}_{VC}^*: \mathbf{Y} \in \mathbb{C}^{N_1 \times 2N_2 N_3} \to \mathbf{X} \in \mathbb{C}^{(N_1+N_2-1) \times N_3}$$

$$\mathcal{H}_{VC}^* \mathbf{Y} = \left[ \frac{\mathcal{H}^*\mathbf{Y}_1 + (\mathcal{H}^*\mathbf{Y}_{1+N_3})^\dagger}{2}, \cdots, \frac{\mathcal{H}^*\mathbf{Y}_{N_3} + (\mathcal{H}^*\mathbf{Y}_{2N_3})^\dagger}{2} \right], \quad (5)$$

where $\mathbf{Y}_i$ denotes the submatrix from the $((i\text{-}1) \times N_2 + 1)^{\text{th}}$ to the $(i \times N_2)^{\text{th}}$ column of the matrix.

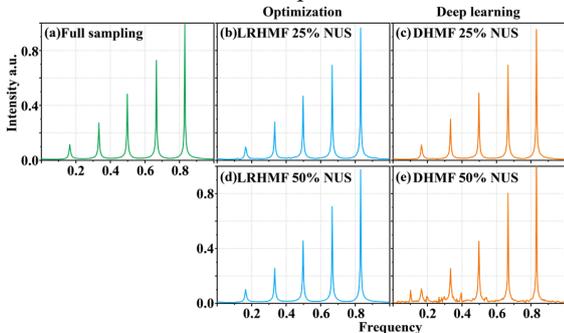

Fig. 1. Reconstructed spectra with mismatch in sampling rates. (a) Noise-free full sampled spectrum. Reconstructed spectra with sampling rate of 25% by (b) LRHMF, (c) DHMF trained by dataset with sampling rate of 25%. (d-e) are the same as (b-c) except for reconstructed spectra with sampling rate of 50%.

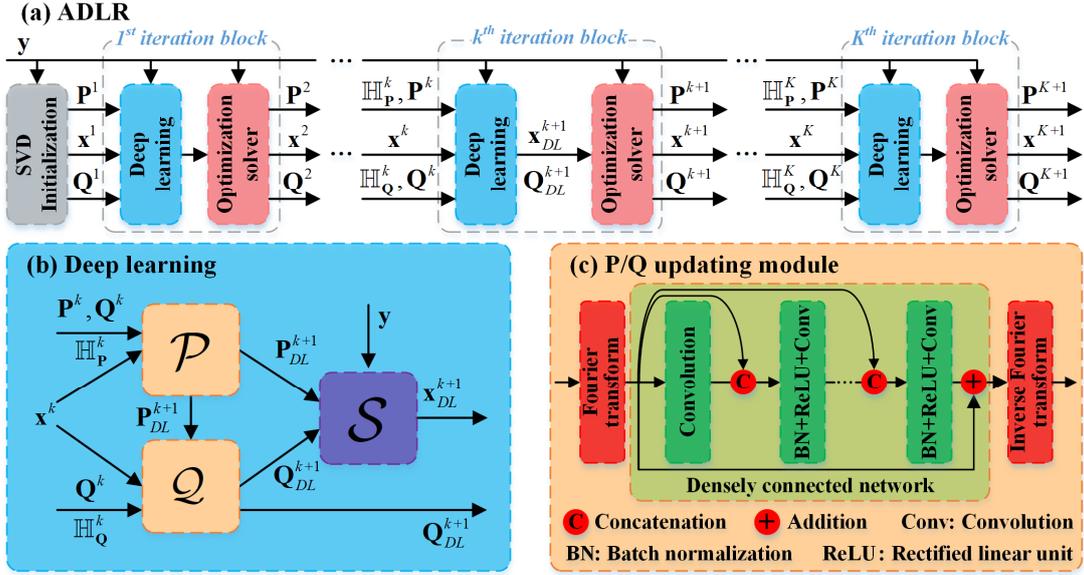

Fig. 2. Overview of ADLR for the exponential function. (a) The iterative structure of ADLR of $K$ subsequent iteration blocks. The detailed structure in (b) the deep learning solver, and (c) the updating modules P and Q.

## B. Low-Rank Hankel Matrix Reconstruction

The minimization of the nuclear norm of Hankel matrix $\mathcal{H}\mathbf{x}$ is introduced into Low-rank Hankel matrix (LRHM) [9] to reconstruct exponential function $\mathbf{x}$:

$$\min_{\mathbf{x}}\|\mathcal{H}\mathbf{x}\|_* + \frac{\lambda}{2}\|\mathbf{y} - \mathcal{U}\mathbf{x}\|_2^2, \qquad (6)$$

where parameter $\lambda$ ($>0$) trades off the nuclear norm and the alignment with the sampled data $\mathbf{y}$. Time-consuming SVD is commonly used to compute the nuclear norm, defined as the summation of the singular values. For the reduction of reconstruction times, the matrix factorization [27] $\mathbf{X} = \mathbf{PQ}^H$ is applied in LRHMF [8] and computes the Frobenius norm instead of the nuclear norm:

$$\|\mathbf{X}\|_* = \min_{\mathbf{P},\mathbf{Q}} \frac{1}{2}(\|\mathbf{P}\|_F^2 + \|\mathbf{Q}\|_F^2) \; s.t. \; \mathbf{X} = \mathbf{PQ}^H. \qquad (7)$$

Then, Eq. (6) is reformulated and solved by alternating direction method of multipliers (ADMM) [34]:

$$\min_{\mathbf{x},\mathbf{P},\mathbf{Q}} \frac{1}{2}(\|\mathbf{P}\|_F^2 + \|\mathbf{Q}\|_F^2) + \frac{\lambda}{2}\|\mathbf{y} - \mathcal{U}\mathbf{x}\|_2^2 \; s.t. \; \mathcal{H}\mathbf{x} = \mathbf{PQ}^H. \qquad (8)$$

## C. Solver with Low-Rank Hankel Matrix

Recently, inspired by deep learning, DHMF [20] unrolls the optimization solver of LRHMF to convolutional neural networks. It consists of several iteration blocks and each block utilizes the densely connected convolutional network [35] as the backbone. As the exponential function proceeds through these blocks, the distribution of singular values approximates that of the full sampling gradually. In the final block, DHMF provides exponential functions with high fidelity and low rank.

However, DHMF fails to reconstruct exponential signals when mismatches occur, especially when the sampling rate of tested signals is higher than that of the training dataset. It is unacceptable that more sampling data provide more valuable information but bring bigger reconstruction errors. Thus, a robust deep-learning method is worth researching.

## III. THE PROPOSED METHOD

In this section, the structure of ADLR is described. Moreover, using MR signals synthesized by exponentials, the intermediate variables in each iteration block are analyzed for network interpretability.

### A. Structure for Exponential Functions

*1) Solver with Low-Rank Hankel Matrix*

Unlike LRHMF solving Eq. (8), the penalty function model [36] is adopted instead of the widely-used augmented Lagrange model. The latter model has four variables due to the added Lagrange multiplier, while the former has only three variables to solve:

$$\min_{\mathbf{x},\mathbf{P},\mathbf{Q}} \frac{1}{2}(\|\mathbf{P}\|_F^2 + \|\mathbf{Q}\|_F^2) + \frac{\lambda}{2}\|\mathbf{y} - \mathcal{U}\mathbf{x}\|_2^2 + \frac{\beta}{2}\|\mathcal{H}\mathbf{x} - \mathbf{PQ}^H\|_F^2, \qquad (9)$$

where $\beta > 0$ is a penalty coefficient. The detailed introduction of the solver is described in supplementary material S5. The solver in the $k^{\text{th}}$ iteration is as follows:

$$\begin{cases} \mathbf{P}^{k+1} = \beta\mathcal{H}\mathbf{x}^k \mathbf{Q}^k \left(\beta(\mathbf{Q}^k)^H \mathbf{Q}^k + \mathbf{I}\right)^{-1} \\ \mathbf{Q}^{k+1} = \beta(\mathcal{H}\mathbf{x}^k)^H \mathbf{P}^{k+1} \left(\beta(\mathbf{P}^{k+1})^H \mathbf{P}^{k+1} + \mathbf{I}\right)^{-1} \\ \mathbf{x}^{k+1} = (\lambda\mathcal{U}^*\mathcal{U} + \beta\mathcal{H}^*\mathcal{H})^{-1}(\lambda\mathcal{U}^*\mathbf{y} + \beta\mathcal{H}^*(\mathbf{P}^{k+1}(\mathbf{Q}^{k+1})^H)) \end{cases} \qquad (10)$$

*2) Alternating Deep Low-Rank Approach*

To accelerate the reconstruction procedure of exponential function $\mathbf{x} \in \mathbb{C}^{2N_1+1}$ and increase the reconstruction ability with noise, deep learning is inserted between each optimization solver in Eq. (10). The intermediate variables $\mathbf{P}$, $\mathbf{Q}$ and $\mathbf{x}$ are updated alternately by deep learning and optimization solver with the low-rank prior (Fig. 2(a)).

To match the updated variables in Eq. (10), deep learning (Fig. 2(b)) imitates the iteration step to learn mappings between variables, which consists of the updating module P, the updating module Q and the data consistency module. These three modules and one optimization solver are referred to as one block of the proposed ADLR of the total *K* blocks. The detailed structures in the $k^{\text{th}}$ block, where deep learning updates the three variables marked as $\mathbf{P}_{DL}^{k+1}$, $\mathbf{Q}_{DL}^{k+1}$, $\mathbf{x}_{DL}^{k+1}$ and optimization solver updates the other three variables marked as $\mathbf{P}^{k+1}$, $\mathbf{Q}^{k+1}$, $\mathbf{x}^{k+1}$, are described as follows:

1) *Updating Module P and Q*: The two modules for updating variable $\mathbf{P}_{DL}^{k+1}$ and $\mathbf{Q}_{DL}^{k+1}$ are the same structure, except for the input and output. Taking the updating module P as an example, one densely connected convolutional network [35], including 2D convolutional layers, the rectified linear unit [37] and batch normalization [38], is selected as the backbone and learns the mapping $\mathcal{P}$ from the input variables $\mathcal{H}\mathbf{x}^k\mathbf{Q}^k$ and $\mathbf{Q}^k$ to output variable $\mathbf{P}_{DL}^{k+1}$. The initial value of variable $\mathbf{P}$ and $\mathbf{Q}$ is generated by SVD with $\mathbf{x}^1 = \mathcal{U}^*\mathbf{y}$, $\mathbf{U},\mathbf{\Sigma},\mathbf{V}=\text{SVD}(\mathcal{H}\mathbf{x}^1)$, $\mathbf{P}^0=\mathbf{U}\mathbf{\Sigma}^{0.5}$, $\mathbf{Q}^0=\mathbf{V}^H\mathbf{\Sigma}^{0.5}$. To utilize the low-rank prior in exponential functions, only the first few columns of variable $\mathbf{P}^0$ and $\mathbf{Q}^0$, denoted as $\mathbf{P}^1 \in \mathbb{C}^{N_1 \times R}$ and $\mathbf{Q}^1 \in \mathbb{C}^{N_1 \times R}$, respectively, are inputted into the updating module P.

Considering the variables $\mathbf{P}^k$ and $\mathbf{P}_{DL}^k$ in all the K blocks are the intermediate values of the solved variable $\mathbf{P}$, the variable set $\mathbb{H}_{\mathbf{P}}^k = \{\mathbf{P}^1, \mathbf{P}_{DL}^i, \mathbf{P}^i | i=2,...,k\}$ if $k \geq 2$ else $\mathbb{H}_{\mathbf{P}}^k = \{\mathbf{P}^1\}$, called historical information of $\mathbf{P}_{DL}^{k+1}$, is added to the input of modules. Similar to DHMF [20], Fourier transform and inverse Fourier transform are applied before and after the densely connected network. Residual connections [39] are employed to improve the learning ability of modules (Fig .2(c)). Finally, the mapping $\mathcal{P}$ of the updating module P in the $k^{\text{th}}$ block can be expressed:

$$\mathbf{P}_{DL}^{k+1} = \mathcal{P}(\mathcal{H}\mathbf{x}^k\mathbf{Q}^k, \mathbf{Q}^k, \mathbb{H}_{\mathbf{P}}^k) + \mathbf{P}^k. \quad (11)$$

Different from the updating module P, the input of the updating module Q in $k^{\text{th}}$ block includes $(\mathcal{H}\mathbf{x}^k)^H\mathbf{P}_{DL}^{k+1}$, $\mathbf{P}_{DL}^{k+1}$ and variable set $\mathbb{H}_{\mathbf{Q}}^k = \{\mathbf{Q}^1, \mathbf{Q}_{DL}^i, \mathbf{Q}^i | i=2,...,k\}$ if $k \geq 2$ else $\mathbb{H}_{\mathbf{Q}}^k = \{\mathbf{Q}^1\}$, and the output is $\mathbf{Q}_{DL}^{k+1}$:

$$\mathbf{Q}_{DL}^{k+1} = \mathcal{Q}((\mathcal{H}\mathbf{x}^k)^H\mathbf{P}_{DL}^{k+1}, \mathbf{P}_{DL}^{k+1}, \mathbb{H}_{\mathbf{Q}}^k) + \mathbf{Q}^k. \quad (12)$$

2) *Data Consistency Module*: The output $\mathbf{x}_{DL}^{k+1}$ of the module is calculated without the learnable convolutional network:

$$\mathbf{x}_{DL}^{k+1} = (\gamma_{DL}^k \mathcal{U}^*\mathcal{U} + \mathcal{H}^*\mathcal{H})^{-1}(\gamma_{DL}^k \mathcal{U}^*\mathbf{y} + \mathcal{H}^*(\mathbf{P}_{DL}^{k+1}(\mathbf{Q}_{DL}^{k+1})^H)), \quad (13)$$

Equation (13) can be reformulated as:

$$(\mathbf{x}_{DL}^{k+1})_n = \mathcal{S}(\mathcal{U}^*\mathbf{y}, \tilde{\mathbf{x}}_{DL}^{k+1}, \gamma_{DL}^k) = \begin{cases} \frac{\gamma_{DL}^k (\mathcal{U}^*\mathbf{y})_n + (\tilde{\mathbf{x}}_{DL}^{k+1})_n}{1+\gamma_{DL}^k}, & if\ n \in \Omega \\ (\tilde{\mathbf{x}}_{DL}^{k+1})_n, & if\ n \notin \Omega \end{cases}, \quad (14)$$

where $\tilde{\mathbf{x}}_{DL}^{k+1} = \mathcal{H}^*(\mathbf{P}_{DL}^{k+1}(\mathbf{Q}_{DL}^{k+1})^H)$ is the reconstructed signal before data consistency, $\gamma_{DL}^k = \frac{\lambda_{DL}^k}{\beta_{DL}^k}$ is a learnable parameter that balances the reliability between $\tilde{\mathbf{x}}_{DL}^{k+1}$ and sampling signal $\mathcal{U}^*\mathbf{y}$ in the set $\Omega$ of sampling position. This module forces the output $\mathbf{x}_{DL}^{k+1}$ to be aligned to the undersampled signal [19, 20].

3) *Optimization Solver*: To match the output variables of the above three modules and explicitly introduce the low-rank property, the solver in Eq. (10) follows the deep learning of ADLR and generates the output variable $\mathbf{P}^{k+1}$, $\mathbf{Q}^{k+1}$, $\mathbf{x}^{k+1}$ of the $k^{\text{th}}$ iteration block with the learnable parameter $\gamma^k$. The regularization parameter $\beta^k$ is also learnable during updating variables $\mathbf{P}$ and $\mathbf{Q}$, denoted as $\beta_{\mathbf{P}}^k$ and $\beta_{\mathbf{Q}}^k$ respectively.

Thus, the whole procedure of ADLR in $k^{\text{th}}$ iteration block is expressed as:

$$\begin{cases} \mathbf{P}_{DL}^{k+1} = \mathcal{P}(\mathcal{H}\mathbf{x}^k\mathbf{Q}^k, \mathbf{Q}^k, \mathbb{H}_{\mathbf{P}}^k) + \mathbf{P}^k \\ \mathbf{Q}_{DL}^{k+1} = \mathcal{Q}((\mathcal{H}\mathbf{x}^k)^H\mathbf{P}_{DL}^{k+1}, \mathbf{P}_{DL}^{k+1}, \mathbb{H}_{\mathbf{Q}}^k) + \mathbf{Q}^k \\ \mathbf{x}_{DL}^{k+1} = \mathcal{S}\left(\mathcal{U}^*\mathbf{y}, \mathcal{H}^*(\mathbf{P}_{DL}^{k+1}(\mathbf{Q}_{DL}^{k+1})^H), \gamma_{DL}^k\right) \\ \mathbf{P}^{k+1} = \beta_{\mathbf{P}}^k(\mathcal{H}\mathbf{x}_{DL}^{k+1})\mathbf{Q}_{DL}^{k+1}\left(\beta(\mathbf{Q}_{DL}^{k+1})^H\mathbf{Q}_{DL}^{k+1}+\mathbf{I}\right)^{-1} \\ \mathbf{Q}^{k+1} = \beta_{\mathbf{Q}}^k(\mathcal{H}\mathbf{x}_{DL}^{k+1})^H\mathbf{P}^{k+1}\left(\beta(\mathbf{P}^{k+1})^H\mathbf{P}^{k+1}+\mathbf{I}\right)^{-1} \\ \mathbf{x}^{k+1} = \mathcal{S}\left(\mathcal{U}^*\mathbf{y}, \mathcal{H}^*\left(\mathbf{P}^{k+1}(\mathbf{Q}^{k+1})^H\right), \gamma^k\right) \end{cases} \quad (15)$$

3) *Loss Function and Hyperparameter*

As a common differentiable metric to compute the difference between the reconstructed and the full sampling, the mean squared error (MSE) is selected as the loss function in the ADLR. Considering the data consistency module forcing the fidelity of the reconstructed signal in sampling positions without learning, it weakens the ability of the direct guide by fully sampled signals for the learned signal $\mathcal{H}^*(\mathbf{P}^{k+1}(\mathbf{Q}^{k+1})^H)$ before data consistency module. For effective guidance to learnable parameters $\Theta$, the MSE between $\mathcal{H}^*(\mathbf{P}^{k+1}(\mathbf{Q}^{k+1})^H)$ and fully sampled signal $\mathbf{x}$ is added. Thus, the loss function for variables in the optimization solver of the $k^{\text{th}}$ block is defined as $\mathcal{L}(\Theta)^k = \|\mathbf{x}^{k+1}-\mathbf{x}\|_2^2 + \alpha\|\mathcal{H}^*(\mathbf{P}^{k+1}(\mathbf{Q}^{k+1})^H)-\mathbf{x}\|_2^2$, where the trade-off parameter $\alpha$ is set as $10^{-2}$ to balance the guidance of loss before and after data consistency module. Similarly, the loss function of deep learning solvers in the $k^{\text{th}}$ block is defined as $\mathcal{L}(\Theta)_{DL}^k = \|\mathbf{x}_{DL}^{k+1}-\mathbf{x}\|_2^2 + \alpha\|\mathcal{H}^*(\mathbf{P}_{DL}^{k+1}(\mathbf{Q}_{DL}^{k+1})^H)-\mathbf{x}\|_2^2$. In a word, the whole loss function $\mathcal{L}(\Theta)$ with all the learnable parameters $\Theta$ in the *K* blocks is written as:

$$\mathcal{L}(\Theta) = \sum_{k=1}^K \mathcal{L}(\Theta)_{DL}^k + \mathcal{L}(\Theta)^k. \quad (16)$$

The proposed ADLR has 10 blocks in total and each densely connected convolutional network in the block consists of 6 layers. Except for the last layer with 2 filters to output the real and imaginary parts of complex signals, each layer includes 12 filters with kernel size 3×3. The parameter *R* is usually twice the maximum rank (10) of fully sampled signals, which explicitly constrains the property of signals with low rank.

Under a batch size of 40, all the learnable parameters, including convolutional kernels and all the regularization parameters $\lambda, \beta$, are updated by Adam optimizer[40]. The learning rate is initially set to $10^{-3}$ and decrease by the multiplicative factor of $10^{-0.5}$ when the reconstruction error in the validation set stops decreasing. The error is defined as the relative $\ell_2$ norm error (RLNE) between the fully sampled signal $\mathbf{x}$ and the reconstructed signal $\hat{\mathbf{x}}$: $\text{RLNE}(\mathbf{x},\hat{\mathbf{x}}) = \frac{\|\mathbf{x}-\hat{\mathbf{x}}\|_2}{\|\mathbf{x}\|_2}$. Once the learning rate reaches to $10^{-4.5}$, the training procedure is terminated.

### B. Structure for MRI

For a multi-coils 2D MRI k-space signal $\mathcal{K} \in \mathbb{C}^{A \times Z \times C}$, where A, Z and C denote the length of frequency encoding, phase encoding and number of coils, respectively, inspired by 1D scheme [13], the whole signal is split into *A* rows of k-space signal $\mathbf{K} \in \mathbb{C}^{Z \times C}$ by inverse Fourier transform on frequency encoding dimension. Owing to the low rankness originated from the correlation between coils [13, 14, 41], similar to Eq. (9), the Hankel form $\mathcal{H}_{VC}\mathbf{K}$ of k-space signal $\mathbf{K}$ with virtual coil [33] is utilized in the reconstruction problem of multi-coils MRI from the undersampled k-space signal $\mathbf{Y}$:

$$\min_{\mathbf{K},\mathbf{P},\mathbf{Q}} \frac{1}{2}(\|\mathbf{P}\|_F^2+\|\mathbf{Q}\|_F^2)+\frac{\lambda}{2}\|\text{vec}(\mathbf{Y}-\mathcal{U}\mathbf{K})\|_2^2+\frac{\beta}{2}\|\mathcal{H}_{VC}\mathbf{K}-\mathbf{PQ}^H\|_F^2, \quad (17)$$

where $\text{vec}(\cdot)$ means vectorization of the matrix.

The detailed structure of ADLR for MRI reconstruction (Fig. 3) in the $k^{\text{th}}$ block is expressed as follows:

$$\begin{cases} \mathbf{P}_{DL}^{k+1}=\mathcal{P}(\mathcal{H}_{VC}\mathbf{K}^k\mathbf{Q}^k, \mathbf{Q}^k, \mathbb{H}_\mathbf{P}^k)+\mathbf{P}^k \\ \mathbf{Q}_{DL}^{k+1}=\mathcal{Q}((\mathcal{H}_{VC}\mathbf{K}^k)^H\mathbf{P}_{DL}^{k+1}, \mathbf{P}_{DL}^{k+1}, \mathbb{H}_\mathbf{Q}^k)+\mathbf{Q}^k \\ \mathbf{K}_{DL}^{k+1}=\mathcal{S}\left(\mathcal{U}^*\mathbf{Y}, \mathcal{H}_{VC}^*(\mathbf{P}_{DL}^{k+1}(\mathbf{Q}_{DL}^{k+1})^H), \frac{\beta_{DL}^k}{\lambda_{DL}^k}\right) \\ \mathbf{P}^{k+1}=\beta^k(\mathcal{H}_{VC}\mathbf{K}_{DL}^{k+1})\mathbf{Q}_{DL}^{k+1}(\beta(\mathbf{Q}_{DL}^{k+1})^H\mathbf{Q}_{DL}^{k+1}+\mathbf{I})^{-1} \\ \mathbf{Q}^{k+1}=\beta^k(\mathcal{H}_{VC}\mathbf{K}_{DL}^{k+1})^H\mathbf{P}^{k+1}(\beta(\mathbf{P}^{k+1})^H\mathbf{P}^{k+1}+\mathbf{I})^{-1} \\ \mathbf{K}^{k+1}=\mathcal{S}\left(\mathcal{U}^*\mathbf{Y}, \mathcal{H}_{VC}^*(\mathbf{P}^{k+1}(\mathbf{Q}^{k+1})^H), \frac{\beta^k}{\lambda^k}\right) \end{cases} \quad (18)$$

Initial values of variables $\mathbf{P}$ and $\mathbf{Q}$ are generated by SVD that $\mathbf{X}^1=\mathcal{U}^*\mathbf{Y}$, $\mathbf{U},\mathbf{\Sigma},\mathbf{V}=\text{SVD}(\mathcal{H}_{VC}^*\mathbf{X}^1)$, $\mathbf{P}^0=\mathbf{U}\mathbf{\Sigma}^{0.5}$, $\mathbf{Q}^0=\mathbf{V}^H\mathbf{\Sigma}^{0.5}$ and the first $R$ columns are selected, denoted as $\mathbf{P}^1 \in \mathbb{C}^{A \times R}$, $\mathbf{Q}^1 \in \mathbb{C}^{A \times R}$. Here, $R$ is set as 40 due to the higher rank of MRI k-space data than NMR spectra.

Considering that magnitude images are widely used in clinical diagnosis, the Frobenius norm of error of the square root of the sum of squares between fully sampled and reconstructed multi-coil signals is added to the loss function. To save the memory in training, we reduce the batch size and the number $K$ of iteration blocks to 32 and 5, respectively.

For simplicity, the structure of ADLR for MRI is still denoted as ADLR.

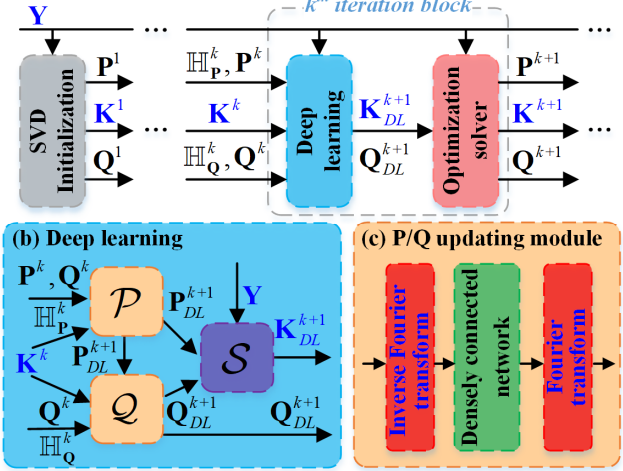

Fig. 3. The overview of ADLR for MRI image reconstruction. (a) The structure of ADLR of K subsequent iteration blocks. The detailed structure in (b) the deep learning solver, and (c) the updating modules P and Q.

### C. Network Interpretability

To illustrate the performance of the network of the ADLR, the MR signal is synthesized by an exponential function with five peaks according to Eq.(1), whose generation parameters are listed in supplementary material S1. Besides, the RLNE and the nuclear norm of the reconstructed signal $\mathbf{x}$ in all 10 blocks are analyzed (Fig. 3).

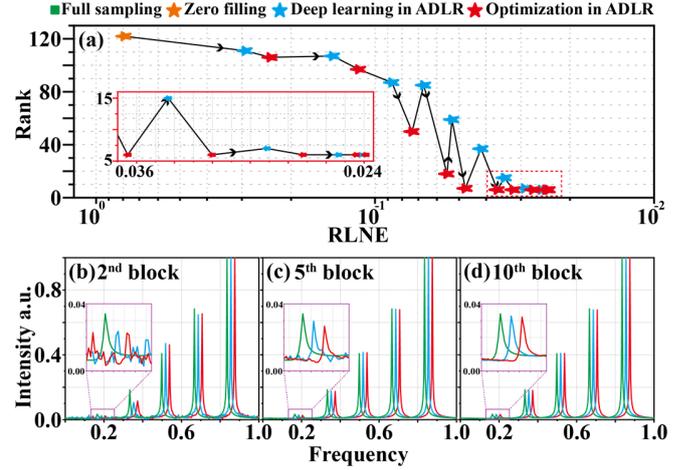

Fig. 4. The changing trends of the reconstructed signal over iterations. (a) The RLNE and rank of the reconstructed signal. (b-d) The spectra of the full sampling and the reconstructed signals in deep learning, optimization solver of the $2^{\text{nd}}$, $5^{\text{th}}$ and $10^{\text{th}}$ blocks.

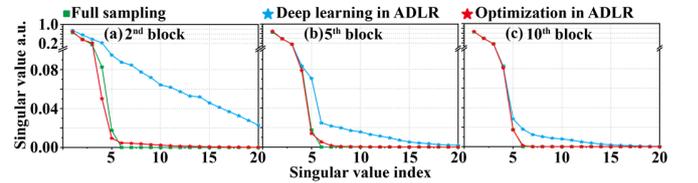

Fig. 5. The singular values of the full sampling and the intermediate variables $\mathbf{PQ}^H$ in deep learning, optimization solver of (a) the $2^{\text{nd}}$ blocks, (b) the $5^{\text{th}}$ blocks and (c) the $10^{\text{th}}$ blocks.

For the noisy synthetic exponential with Gaussian noise of standard deviation 0.03, the deep learning solvers reconstruct the signal with a lower RLNE but a higher rank than their input signal in some blocks (Fig. 4(a)). On the contrary, the optimization solvers can effectively reduce the rank of signals, close to the true rank 5 (of noise-free fully sampled signal). As the signal passed by more sequential blocks, the solution with low RLNE is gradually obtained, which reflects the high fidelity of the reconstructed spectra, especially for the peak of low intensity (Fig. 4(b-d)). In the last block, the reconstructed signal achieves a low reconstruction error with low-rank property, almost the same as the noise-free full sampling.

The P and Q modules in deep learning solvers are not consistent with optimization solvers but the difference enlarges the model capacity in the reconstruction of noisy signals. In the deep learning solver, the reconstructed signal is expected to be low RLNE. Besides, due to the loss function that minimizes the reconstruction errors of each solver, the variables $\mathbf{P}$ and $\mathbf{Q}$ are expected to achieve a reconstructed signal close to the noiseless full sampling after a low-rank optimization solver. The signal of deep learning solvers does not focus on low rank and even increases the rank (Fig. 4(a)). In the optimization solver, the Frobenius norms of the variables $\mathbf{P}$ and $\mathbf{Q}$ are minimized. According to $\|\mathbf{X}\|_*=\min_{\mathbf{P},\mathbf{Q}}\frac{1}{2}(\|\mathbf{P}\|_F^2+\|\mathbf{Q}\|_F^2)$ $s.t. \mathbf{X}=\mathbf{PQ}^H$, it leads to a reduction in the nuclear norm of the reconstructed

signals (Fig. 5), which is a convex relaxation of the matrix rank [42]. Thus, the optimization solver compensates for the disadvantage that deep learning solvers increase the rank.

*D. Synthetic Exponential Function Reconstructions*

To demonstrate the robust reconstruction ability of ADLR, the signal in Fig. 3 is undersampled with different sampling patterns and sampling rates from 10% to 50%. Gaussian noise of standard deviation 0.03 is additive. The optimization-based compressed sensing (CS) [22], LRHMF, and deep-learning-based FID-Net [31] and DHMF are compared. The regularization parameters of optimization methods are chosen according to the lowest RLNE. The deep learning methods are trained only once with exponentials at a sampling rate of 25%, whose detailed training scheme is described in Section IV.A.

At the sampling rate of 50%, CS fails to reconstruct the signal (Fig. 6(b)) because the sparsity of decaying exponentials in the frequency domain is not satisfied well. FID-Net can recover the high-intensity peaks but not the low-intensity ones whose intensity is less than 0.04 (Fig. 6(d)). Intuitively, more sampled data provide more information for the reconstructed signal and can decrease reconstruction error. However, DHMF [20] fails to reconstruct the exponential signal with a sampling rate of 50% when trained in the dataset with a sampling rate of 25% (Fig. 6(e)). For comparison, ADLR can still effectively reconstruct the signal, especially the low-intensity peak (Fig. 6(f)). The result is similar to that obtained from optimization-based LRHMF [8] (Fig. 6(c)) and indicates ADLR can reconstruct the signal well when sampling rates mismatch.

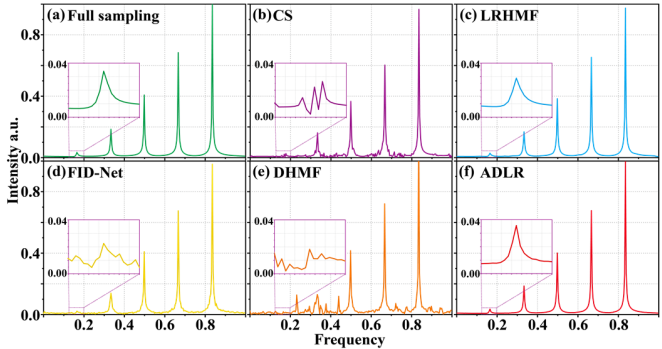

Fig. 6. Reconstructions of the synthetic exponential with additive Gaussian noise of standard deviation 0.03. (a) The spectrum of the noise-free fully sampled exponential. Spectra reconstructed from 50% of fully-sampled data by (b) CS, (c) LRHMF, and deep learning methods (d) FID-Net, (e) DHMF and (d) ADLR trained with sampling rate 25%.

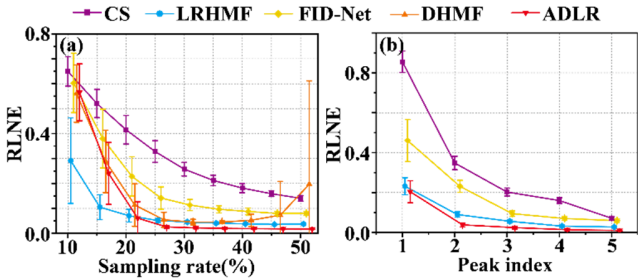

Fig. 7. Mean and standard deviation of reconstruction errors of synthetic spectra. (a) RLNEs of spectra reconstructed from 10% to 50% data. (b) RLNEs of each peak of spectra reconstructed from 50% data. The RLNE of the 1st peak of DHMF is large (1.2±2.3) and not shown. RLNEs are computed over 100 sampling trials and shifted along X axis for visual convenience.

Furthermore, the RLNEs of each spectral peak are also calculated. The separation points of adjacent peaks is determined by the minimum between them in the noise-free full sampling. Under 100 resamples at a sampling rate of 50% (Fig. 7(b)), the proposed ADLR achieves the lowest mean RLNE of all peaks in all methods. For the whole signals, ADLR still achieves the lowest mean RLNE under multiple sampling rates from 20% to 50% (Fig. 7(a)). Besides, the RLNE of DHMF increases after the sampling rate reaches 35% but ADLR not.

## IV. EXPERIMENTAL RESULTS

Here, we describe the training scheme and reconstruction experiments of NMR spectroscopy and multi-coils MRI with mismatches, respectively.

*A. NMR Spectroscopy Reconstructions*

*1) Reconstruction scheme for NMR*

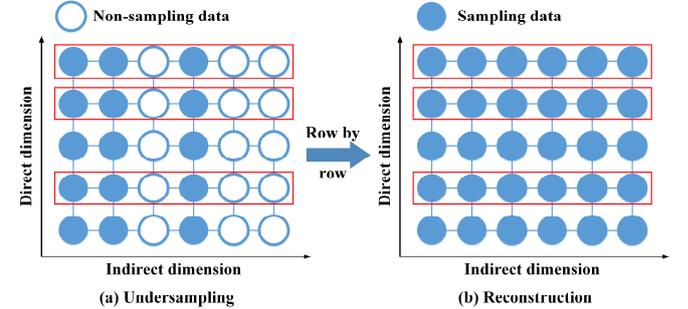

Fig. 8. The schematic diagram of reconstructing the undersampled 2D NMR spectroscopy. (a) Undersampling operator in the indirect dimension. (b) Reconstructing the NMR spectrum row by row.

Biological NMR spectroscopy is one of the important analysis devices for modern chemistry [43], biology [44] and life science [18]. For the 2D NMR spectroscopy, the two dimensions are named direct dimension and indirect dimension. Time-domain signals along with the direct dimension can be represented by the summation of damped exponential components [7]. Owing to the sampling mechanism of NMR spectroscopy (Fig. 8), the undersampling operator only occurs in the indirect dimension. As a result, reconstructing 2D NMR spectroscopy can be represented as the reconstruction of 1D exponential functions row by row.

*2) Training Dataset*

The total 40000 fully sampled exponential functions **x** of length 255 with sampling interval $\Delta t=1$ are simulated (Fig. 9) for the training of ADLR according to Eq. (1). To test the robustness to noise, the additive Gaussian noise **n** with standard deviation $\sigma$ selected uniformly from 0 to 0.04 is employed. All the generation parameters are listed in Table I.

To train the ADLR for reconstruction, the undersampled signal **y** is obtained from the exponential function **x** by Poisson-gap sampling scheme $\mathcal{U}$[45] defined as **y**=$\mathcal{U}$**x**+**n**. To rich the diversity of undersampled signals and increase the robustness for various sampling positions, all the sampling operators $\mathcal{U}$ with the same sampling rate are different in sampling position. In the end, 40000 pairs of training data $(\mathbf{y}_t, \mathcal{U}_t, \mathbf{x}_t)$ are generated, where $t=1,2,3,...,4000$ means the $t$-th sampling trial. Here, 90% (36000) of data pairs are selected to update learnable parameters and the others 10% (4000) are employed to evaluate the performance of learned parameters and pick the optimal hyperparameters.

TABLE I
PARAMETER OF THE TRAINING DATASET

| Parameters | Minimum | Increment | Maximum |
|---|---|---|---|
| Number of peak $G$ | 1 | 1 | 10 |
| Frequency $f_g$ | 0.00 | $2.2\times10^{-16}$ | 1.00 |
| Phase $\phi_g$ | 0.00 | $8.8\times10^{-16}$ | $2\pi$ |
| Amplitude $A_g$ | 0.05 | $2.2\times10^{-16}$ | 1.00 |
| Decay factor $\tau_g$ | 10.00 | $2.8\times10^{-14}$ | 179.20 |
| Standard deviation $\sigma$ | 0.00 | $6.9\times10^{-18}$ | 0.04 |

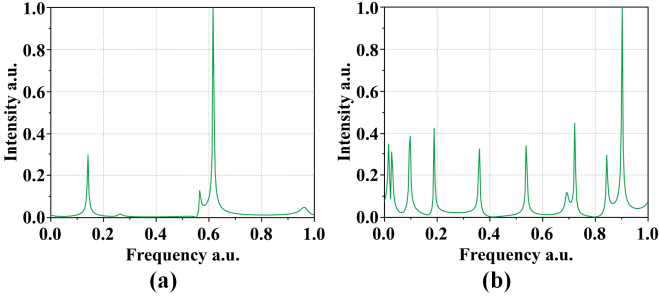

Fig. 9. Spectra of the synthetic exponentials in the training dataset. (a) the synthetic exponential with 5 peaks; (b) the synthetic exponential with 10 peaks.

*3) Reconstruction Experiments*

Due to the limited size of memory in NMR spectrometers, it is hard to load massive different weights of models in a spectrometer for spectrum reconstruction at different sampling rates. Thus, the deep learning method should have the ability to reconstruct spectra successfully, similar to the optimization-based methods, e.g. LRHMF, especially when the sampling rate is higher than that in training. To evaluate the robustness of reconstructing NMR spectra at different sampling rates, all reconstruction methods for exponential functions are compared.

As an example, the 2D $^1$H-$^{15}$N TROSY spectra of Ubiquitin are reconstructed from 25% and 50% of the fully sampled data by retrospective undersampling (Fig. 10). To evaluate the quality of reconstruction, considering the peak regions of fully sampled NMR spectra are focused and other regions mainly contain noise, the square of Pearson correlation coefficient $r^2$ between the peak intensities of the full sampling and the reconstructed spectrum, instead of RLNE, measures the reconstruction quality:

$$r^2(\mathbf{c},\mathbf{d}) = \left(\frac{\sum_{m=1}^{M}(\mathbf{c}_i-\overline{\mathbf{c}})(\mathbf{d}_i-\overline{\mathbf{d}})}{\sqrt{\sum_{m=1}^{M}(\mathbf{c}_i-\overline{\mathbf{c}})^2}\sqrt{\sum_{m=1}^{M}(\mathbf{d}_i-\overline{\mathbf{d}})^2}}\right)^2, \quad (19)$$

where $\mathbf{c}$ and $\mathbf{d}$ are the total intensity of $M$ peaks of fully sampled and reconstructed spectra. $\overline{\mathbf{c}}$ and $\overline{\mathbf{d}}$ are their corresponding mean of intensities. The closer to 1 $r^2$ is, the better reconstruction performance is.

When the sampling rates of training and reconstruction match both at 25%, all the compared methods achieve a high correlation ($r^2>0.99$) except for CS. However, LRHMF generates weak artifacts (black circle in Fig. 10(c)) and FID-Net loses the low-intensity peaks (black arrow in Fig. 10(d)). As the sampling rate of the reconstructed spectra increases to 50%, all methods but DHMF achieve a higher correlation coefficient $r^2$ since DHMF generates strong artifacts (the black circle in Fig. 10(j)). The phenomenon indicates that ADLR can provide the faithfully reconstructed spectrum when sampling rates mismatch. To show the robustness, the other three 2D spectra, i.e. HSQC spectra of Cytosolic CD79b, Gb1 and Ubiquitin, are reconstructed with only part of the fully sampled data. The detailed experiment setups and results are described in supplementary material S2.

### B. MRI Image Reconstructions

As the typical applications of biomedical MR, MRI is a noninvasive and essential tool that visualizes the structures and physiological parameters of organs[46]. However, for deep learning, differences between images of scanned organs, such as the knee and brain, seriously influence the reconstruction quality when the methods trained by the data of one organ reconstruct the data of other organs. Here, ADLR can still effectively alleviate the performance drop in the case mentioned above.

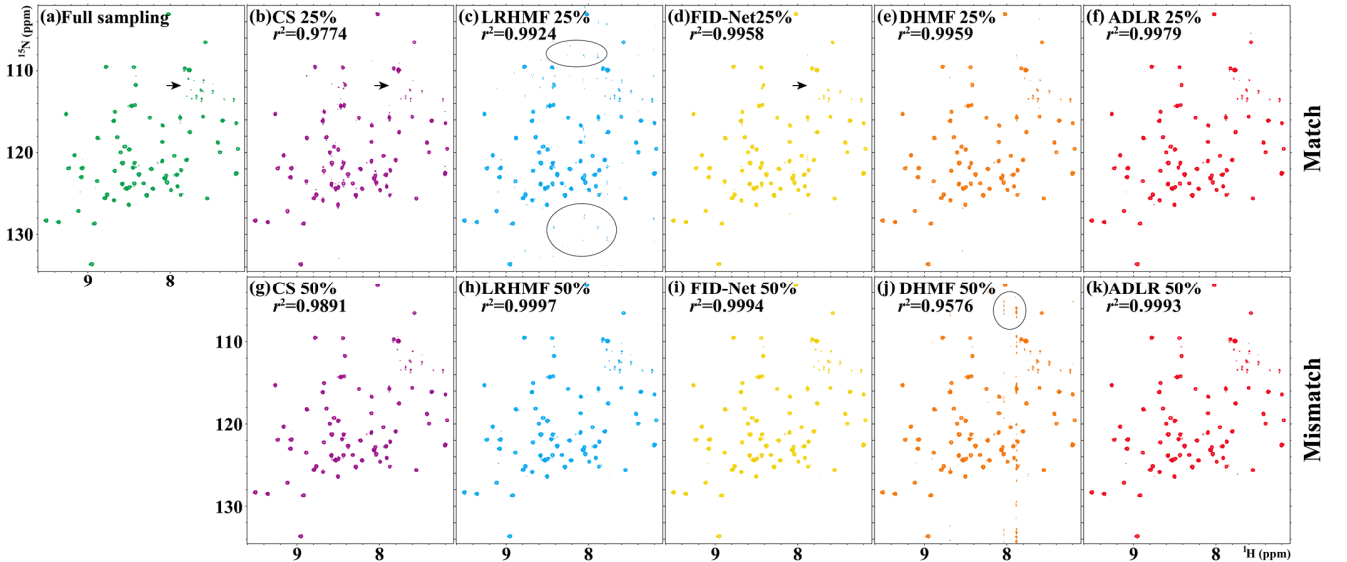

Fig. 10. Reconstructions of 2D $^1$H-$^{15}$N TROSY spectrum of Ubiquitin under the match and mismatch sampling rates. (a) The fully sampled spectrum. The spectra reconstructed by CS, LRHMF, FID-Net, DHMF and ADLR from (b-f) 25% data and (g-k) 50% data, respectively.

*1) Training Dataset*

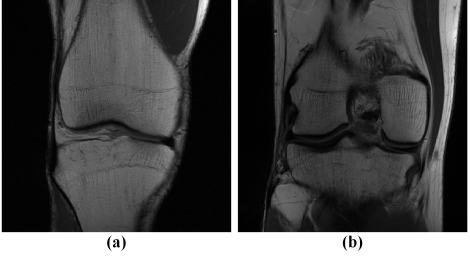

Fig. 11. Magnitude images of the MRI images in the training dataset. (a) and (b) are the 1st and 11th knee MRI images, respectively.

Three multi-coil MRI datasets are used to evaluate the reconstruction robustness of ADLR under different sections and organs, including one coronal and one sagittal proton density weighted (PDw) knee dataset of 15 coils from an open repository [47] and another axial $T_2$ weighted ($T_2$w) brain dataset of 16 coils from the fastMRI [48].

The coronal knee dataset consists of 331 slices from 16 subjects where 273 slices from 13 subjects are used for training (Fig. 11) and the other 58 slices from 3 subjects for test. Similarly, the sagittal knee dataset consists of 245 slices from 16 subjects where 192 slices from 12 subjects are used for training and the other 53 slices from 4 subjects for test. The brain dataset consists of 37 subjects with 10 slices, where 30 subjects are selected for training and 7 subjects for testing.

To save memory and reduce the computation complexity, the number of coils in three datasets is compressed to 4 coils by GCC [49]. Besides, the image sizes of the coronal, sagittal knee and axial brain datasets are center-cropped to 320×314, 320×368 and 320×314, respectively. All the fully sampled MRI data are retrospectively undersampled. For decreasing the sensitivity of changes in data acquisition, different 1D Cartesian undersampling patterns are employed with a sampling rate of 25%. In the training procedure, 1D scheme [13] is adopted for memory efficiency and robustness, where the MRI data is separated into 90% as the training dataset and 10% as the validation set.

*2) Reconstruction Experiments*

To evaluate the robustness of trained ADLR for different organs when reconstructing the multi-coil MRI data, three low-rank-based methods P-LORAKS [50], HDSLR [51] and ODLS [13] are compared. To show the benefit of the optimization solver, the ADLR with only deep learning, denoted as ADLR-D, is also compared. P-LORAKS is based on the optimization and its regularization parameters are chosen according to the lowest reconstruction error (RLNE). The other methods are based on deep learning and trained only once with a coronal PDw knee at a sampling rate of 25%.

When reconstructing the matched coronal knee image and similar sagittal knee image to the training dataset, compared with other methods, the proposed ADLR produces the fewest artifacts and lowest RLNE (Fig. 12(g)(n)). When reconstructing the seriously mismatched axial $T_2$w brain images, due to the lack of robustness, deep-learning methods, HDSLR, ODLS and ADLR-D achieve the high RLNE (>0.12). On the contrary, ADLR achieves the RLNE about 0.10 and produces few artifacts (Fig. 12(u)), as the result of the introduction of data-independent optimization solvers.

To further demonstrate the robustness of ADLR under mismatched reconstruction, methods are trained by the matched dataset. In all methods, ADLR shows the lowest mean of RNLE for all data whether the training and test data are matched or not (Table II). For datasets of sagittal PDw knee (or axial $T_2$w brain), compared to the match reconstruction, the RLNEs of HDSLR, ODLS and ADLR-D in the mismatched reconstruction are increased by 18.78% (or 46.12%), 11.65% (or 35.06%) and 14.96% (or 11.16%), respectively. Notably, ADLR only increasing by 4.91% (or 5.21%) indicates that the alternating combination of the deep learning and the optimization solver can alleviate the deteriorated reconstruction

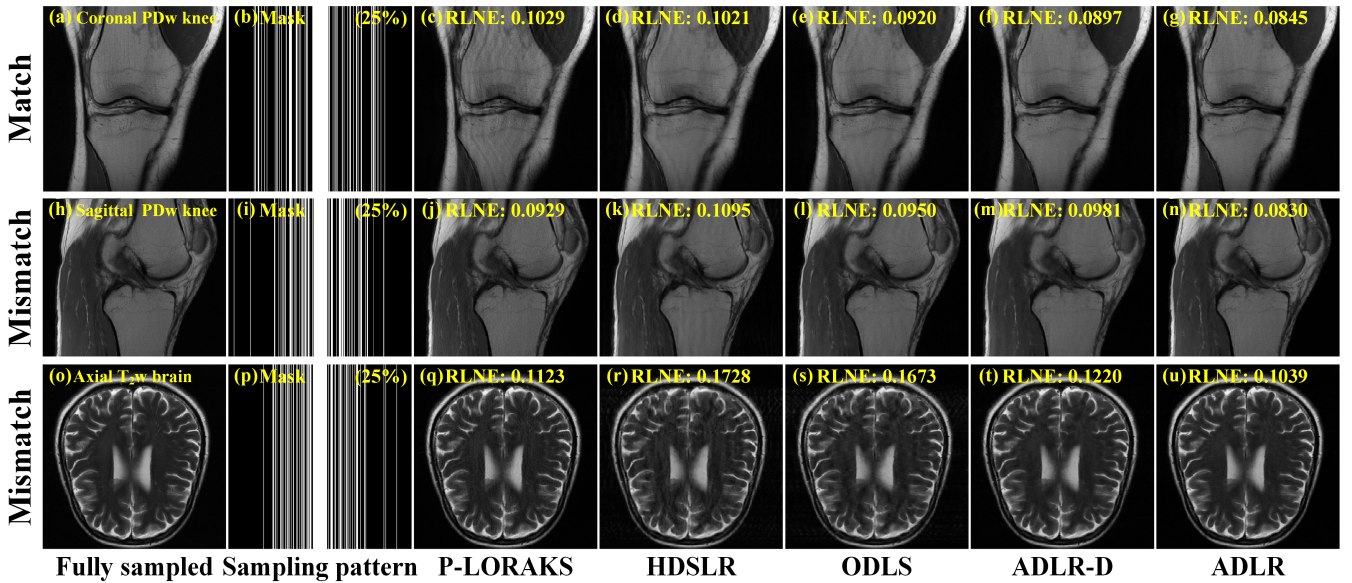

Fig. 12. Reconstruction images of three datasets. (a) The fully sampled images of coronal PDw knee and reconstructed images by (c) P-LORAKS, (d) HDSLR, (e) ODLS, (f) ADLR-D and (g) ADLR with (b) the sampling pattern of 25% data. (h-n) and (o-u) are magnitude images of the full sampling, the corresponding sampling pattern and reconstructed images of sagittal PDw knee and axial $T_2$w brain.

quality due to the mismatched properties between the training and test dataset.

TABLE II
RLNEs (×10⁻²) OF RECONSTRUCTION WITH MATCH AND MISMATCH
[MEAN±STANDARD DEVIATION (DECREMENT OF PERFORMANCE)]

| | P-LORAKS | HDSLR | ODLS | ADLR-D | ADLR |
|---|---|---|---|---|---|
| Coronal knee → Coronal knee | 9.64±1.13 | 11.05±1.65 | 9.61±1.29 | 9.38±1.14 | **8.37±1.10** |
| Coronal knee → Sagittal knee | 8.21±0.82 | 10.24±1.19 (18.78%) | 8.82±1.17 (11.65%) | 9.10±0.95 (14.96%) | **7.84±1.02** (4.91%) |
| PDw knee → T2w brain | 12.79±1.65 | 16.94±1.87 (46.12%) | 16.04±2.40 (35.06%) | 13.28±1.95 (11.16%) | **11.50±1.38** (5.21%) |

Note: The methods reconstruct MRI signals from 25% data. The A→B means the method is trained by dataset A and reconstructs dataset B. The mean and standard deviation are computed over all the reconstructed data. For the dataset B to reconstruct, the percentages in bracket represent the decrement of reconstruction performance, which is computed by the increment of mean of RLNEs under mismatched dataset A divided by the mean of RLNEs under matched dataset B. Here, all the datasets A are from coronal PDw knee. The lowest means of RLNEs are bold faced.

## V. DISCUSSIONS

In this section, we first show the advantage of the alternating structure in ADLR. Then, reconstruction experiments of multi-coil MRI data with different organs and imaging parameters are carried out through a simple modification of ADLR. Besides, the computation time is also compared.

### A. Sequence of Deep Learning and Optimization Solver

To analyze the significance of optimization solvers in ADLR and the alternating structure between deep learning and optimization solvers, three variants of ADLR are trained by the same dataset with a sampling rate of 25%. One is ADLR-D which only includes deep learning. The other two are 10 optimization solvers followed by 10 deep learning structures marked as ADLR-OD, and 10 deep learning structures followed by 10 optimization solvers marked as ADLR-DO.

Because of the different intensities of reconstructed signals, the signals are normalized by their respective nuclear norm. Due to the existence of noise in the signal, we redefine the rank of the signal as the number of singular values greater than $10^{-3}$.

Under a matched sampling rate of 25%, all methods achieve a low reconstruction error (RLNE<0.05). The proposed ADLR obtains rank 6, which is the closest to the true rank 5 (of noise-free fully sampled signal) than other variants (Fig. 13(a-d)). Under a mismatched but a higher sampling rate of 50%, ADLR achieves the lowest RLNE and true rank (Fig. 13(e-f)). However, even with such a higher sampling rate, other variants unexpectedly increase the reconstruction error (Comparing Fig. 13(a-c) with Fig. 13(e-g)). This observation may be explained by their much higher rank than the true rank 5.

The order of optimization and deep learning solvers is important. If the sampling rates of the training and target data are mismatched, sole deep learning (ADLR-D) leads to a much higher reconstruction error and the rank of the solution is much higher than the true one (Comparing Fig. 13(e) and (a)). With several steps of the optimization solver as the initialization (ADLR-OD), the reconstruction error and rank can be reduced (Fig. 13(b) and (f)). But this reduction is marginal when the sampling rate is mismatched since deep learning solvers are hard to adapt to the unseen sampling rate. Thus, the reconstruction error and rank by ADLR-OD is still high in this case. On the contrary, even under the mismatched sampling rate, initializing with deep learning solvers followed by successively optimization solvers (ADLR-DO) enable significant reductions of reconstruction error and rank (Fig. 13(g)). This observation implies that the optimization solver has a very strong ability to improve the robustness since it is generally optimal for the current target data (rather than the averaged optimum for the whole trained database in deep learning).

Compared to the variants, although the deep learning solver of ADLR suffers from mismatched data, fortunately, the subsequent data-independent optimization solver can immediately alleviate the mismatch. Moreover, due to the loss function that minimizes the reconstruction errors of each solver, the deep learning solvers are forced to provide the appropriate initial value of the next optimization solver. Compared with the continuous structure of variants, the alternating structure can alleviate the accumulation of errors from the mismatched data (Fig. 13(h)), leading to the smallest reconstruction error and closest rank to the true one.

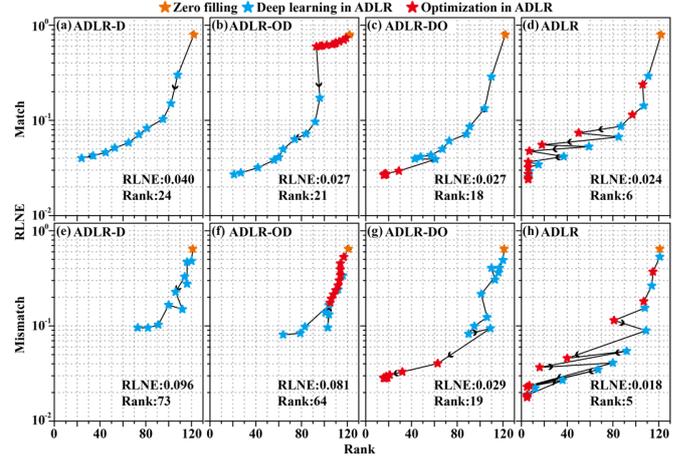

Fig. 13. The rank and RLNE of intermediate signals reconstructed by the ADLR and its three variants. The true rank of the noise-free full sampling is 5 and the added Gaussian noise standard deviation is 0.03. (a) ADLR-D, (b) ADLR-OD, (c) ADLR-DO, and (d) ADLR with the matched sampling rate of 25%. (e-h) The corresponding RLNE and rank under a mismatched sampling rate of 50%.

### B. Robustness of Different Noise Levels and Types

The additive noise, such as under Gaussian distribution and the uniform distribution, can be effectively reduced by ADLR. In the Gaussian noise experiment (Fig. 14(a)), the mean of the noise is 0 and the standard deviation varies from 0.01 to 0.05. For the uniform distribution noise (Fig. 14(b)), the mean of noise is also 0 and the maximum amplitude varies from 0.02 to 0.10. Compared with the traditional optimization method LRHMF, the ADLR trained on the dataset with a sampling rate of 25% can achieve the lower RLNE in all the tested cases under both the sampling rates of 25% and 50%.

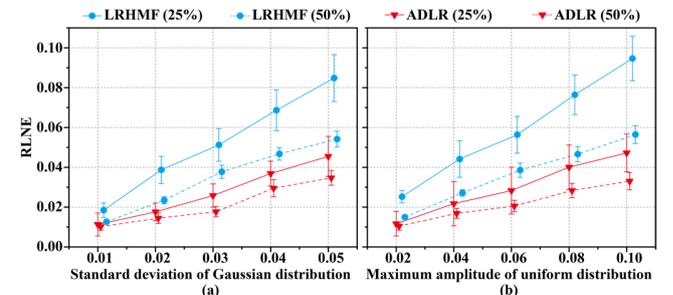

Fig. 14. The mean and standard deviation of reconstruction error of synthetic exponentials with noise of (a) Gaussian distribution and (b) uniform distribution. Under each noise novel, the plot is shifted a little bit for better visualization.

### C. Robustness of Algorithm Parameters

In the proposed ADLR, all regularization parameters $\gamma_{DL}^k$, $\gamma^k$, $\beta_P^k$ and $\beta_Q^k$ are learnable, whose initial value is set as $10^4$, $10^4$, $10^4$ and $10^4$ respectively. The learned values are listed in Tables III and IV for exponential functions and MRI reconstruction. The main non-learnable parameters include the number $K$ of iteration blocks, the number $L$ of layers in each deep learning solver and the weight $\alpha$ in the loss function. Under $K$ from 8 to 12, $L$ from 3 to 7 and $\alpha$ from 0 to $10^{-2}$, all RLNEs achieve about 0.025 from 25% data. Thus, these three parameters are set as 10, 6 and $10^{-2}$ respectively (the black arrows in Fig. 15). Due to the larger size $224 \times 4$ of multi-coil MRI k-space signals than the length 255 of exponential functions in training, the number $K$ of iteration blocks is reduced to 5 for MRI reconstruction and the other non-learnable parameters are the same. The parameters of compared methods are listed in supplementary material S3.

TABLE III
LEARNED REGULARIZATION PARAMETERS OF ADLR FOR THE SYNTHETIC EXPONENTIAL FUNCTIONS AND NMR SPECTROSCOPY

| $k^{th}$ block | 1 | 2 | 3 | 4 | 5 |
|---|---|---|---|---|---|
| $\gamma_{DL}^k$ | 5.2×10⁵ | 6.7×10⁴ | 6.7×10⁴ | 3.7×10¹ | 4.4×10⁰ |
| $\gamma^k$ | 1.9×10⁵ | 1.3×10⁵ | 1.6×10⁰ | 5.4×10⁻¹ | 2.6×10⁻¹ |
| $\beta_P^k$ | 96.0 | 98.4 | 98.5 | 99.5 | 99.0 |
| $\beta_Q^k$ | 87.6 | 97.0 | 98.9 | 95.7 | 102.1 |
| $k^{th}$ block | 6 | 7 | 8 | 9 | 10 |
| $\gamma_{DL}^k$ | 4.0×10⁰ | 2.3×10⁰ | 2.0×10⁰ | 1.2×10⁰ | 4.6×10⁻¹ |
| $\gamma^k$ | 1.4×10⁻¹ | 1.1×10⁻¹ | 9.2×10⁻² | 8.9×10⁻² | 9.5×10⁻² |
| $\beta_P^k$ | 99.2 | 98.5 | 99.4 | 100.0 | 98.4 |
| $\beta_Q^k$ | 102.7 | 105.7 | 105.0 | 107.6 | 109.3 |

TABLE IV
LEARNED REGULARIZATION PARAMETERS OF ADLR FOR THE MULTI-COIL MRI IMAGES

| $k^{th}$ block | 1 | 2 | 3 | 4 | 5 |
|---|---|---|---|---|---|
| $\gamma_{DL}^k$ | 1.4×10⁵ | 4.1×10⁵ | 3.2×10⁵ | 7.7×10³ | 6.4×10⁴ |
| $\gamma^k$ | 2.1×10⁵ | 1.1×10⁵ | 2.4×10⁵ | 2.0×10⁵ | 3.4×10⁵ |
| $\beta_P^k$ | 92.3 | 100.5 | 101.9 | 103.0 | 98.1 |
| $\beta_Q^k$ | 75.7 | 75.4 | 77.7 | 79.2 | 86.0 |

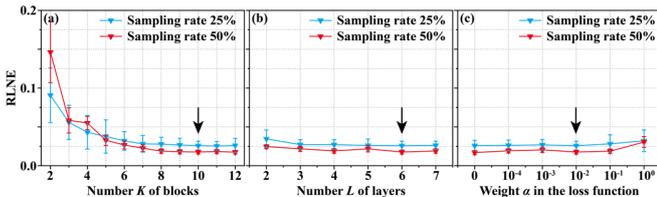

Fig. 15. Reconstruction error of ADLR for exponential functions under sampling rates of 25% and 50% with different (a) numbers $K$ of blocks, (c) numbers $L$ of layers and (c) weights $\alpha$ in the loss function. The black arrows denote the chosen parameters.

### D. Training Strategy

During training, the loss function consists of all values of $\mathcal{L}(\Theta)_{DL}^k$ and $\mathcal{L}(\Theta)^k$ in each deep learning solver and optimization solver, which is a kind of greedy strategy. Compared with the non-greedy strategy that the loss only includes the loss $\mathcal{L}(\Theta)^K$ in the last solver, the greedy strategy can provide a lower reconstruction error.

To better understand, the loss of ADLR is redefined by introducing the weights of loss in each solver:
$$\mathcal{L}(\Theta) = \sum_{k=1}^{K} w_{DL}^k \mathcal{L}(\Theta)_{DL}^k + w^k \mathcal{L}(\Theta)^k, \quad (20)$$
where $w_{DL}^k$ and $w^k$ denote the $k^{th}$ weight of the deep learning solver and the optimization solver respectively. The loss of the greedy strategy is defined when all weights of all solvers $\{w_{DL}^k, w^k | k = 1, ..., K\}$ are set to one, and the loss of non-greedy strategy is defined when only the weight of the last solver $w^K$ is set to one and other weights are zero.

The greedy strategy forces the decrement of loss of each solver (Fig. 16(a)). In contrast, the non-greedy strategy does not effectively decrease the losses before the last solver (Fig. 16(b)), which are at least two orders of magnitude higher than the losses of the greedy strategy. Due to the iterative structure of ADLR, the results of the solver depend on the results of the previous solver. The high losses represent the high deviations of the outputs from the ground truth, thus ultimately leading to the high RLNE of the reconstructed signals.

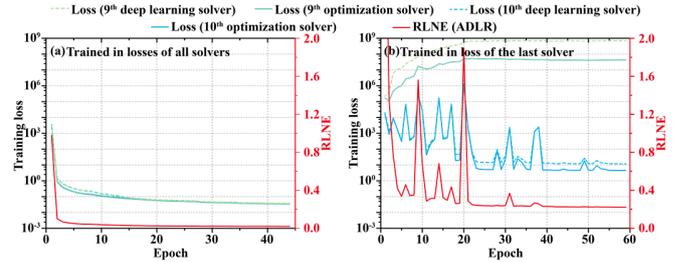

Fig. 16. The losses and reconstruction error of ADLR trained by (a) greedy strategy and (b) non-greedy strategy.

### E. Computation Platform and Time

All the experiments are performed on the computer server whose CPU is a 10-core 3.5GHz Intel i9 9900X with 128-GB RAM and GPU cards are 3 NVIDIA RTX 2080Ti. All deep-learning-based methods are trained and reconstruct signals in one GPU card by using Python 3 and TensorFlow as the backend. The parallel computation by 10 cores is carried out in MATLAB to maximally increase the reconstruction speed of optimization-based methods.

As the deep learning methods are trained, the introduction of optimization solvers slightly increases the training time during one epoch, i.e. the training dataset is iterated once, but fewer epochs are needed to converge than the methods without optimization solvers. The total training time of FID-Net, DHMF and ADLR for exponential functions is about 18, 7 and 8 hours, respectively. Similarly, that of HDSLR, ADLR-D and ADLR for a coronal knee MRI dataset is 1, 11.5 and 9 hours.

During reconstruction, applying the optimization solver in each iteration increases the computation time. However, ADLR is very fast (less than 0.35 seconds) for a signal of size 100×255, due to the adopted fast matrix factorization algorithm [8] and powerful NVIDIA RTX 2080Ti GPU card. In return, the benefit is the improved robustness to sampling rates, achieving the lowest RLNE at a match sampling rate of 25% and a mismatch sampling rate of 50% (Fig. 17).

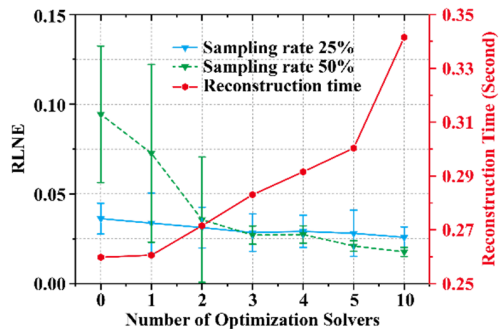

Fig. 17. The mean and standard deviation of reconstruction error and reconstruction time with different numbers of optimization solvers. The size of the reconstructed synthetic signal is 100×255.

For reconstruction of NMR spectra (Table V), compared with CS and FID-Net whose reconstruction times are less than 1 second, ADLR costs time similar to DHMF and computes 10 times faster than LRHMF which needs great amounts of iteration. Besides, ADLR-D and ADLR can reconstruct the multi-coil MRI data in seconds (Table VI), faster than P-LORAKS but slower than other deep learning methods HDSLR and ODLS.

TABLE V
RECONSTRUCTION TIME (SECOND) OF SPECTRA BY DIFFERENT METHODS

| Protein | Type | Size | CS | FID-Net | LRHMF | DHMF | ADLR |
|---|---|---|---|---|---|---|---|
| Gb1 | HSQC | 1466×169 | 0.74 | 0.44 | 22.45 | 2.09 | 2.35 |
| Cytosolic CD79b | HSQC | 116×255 | 0.59 | 0.06 | 3.58 | 0.30 | 0.39 |
| Ubiquitin | HSQC | 576×97 | 0.51 | 0.12 | 4.68 | 0.40 | 0.50 |
| Ubiquitin | TROSY | 512×127 | 0.40 | 0.12 | 5.67 | 0.48 | 0.59 |

Note: The reconstruction time is averaged over four retrospective undersampling with sampling rate of 25%.

TABLE VI
RECONSTRUCTION TIME (SECOND) OF 4-COIL MRI BY DIFFERENT METHODS

| Organ | Plane | Size | P-LORAKS | HDSLR | ODLS | ADLR-D | ADLR |
|---|---|---|---|---|---|---|---|
| Knee | Coronal | 320×314 | 29.15 | 0.12 | 0.10 | 2.45 | 2.70 |
| Knee | Sagittal | 320×368 | 225.19 | 0.10 | 0.11 | 3.62 | 3.94 |
| Brain | Axial | 320×314 | 83.55 | 0.12 | 0.10 | 2.61 | 2.88 |

Note: The reconstruction time is averaged over all the slices in the corresponding test dataset with sampling rate of 25%.

The gap between the reconstruction speeds is mainly caused by the Hankel form of signals and lots of 2D convolutions in the complicated structure. Thus, the combination of efficient optimization solvers and lightweight deep learning structures is the next step in future work. Compared to the Hankel matrix, which increases the dimension of signals and leads to increased computation time, a small Hankel matrix constructed by sliding window [52] effectively increases the speed without obviously sacrificing the quality of reconstruction. In addition, due to the redundancy of exponential signals, signals can be factorized into subspaces and then reconstructed in low dimensions. This strategy has been evidenced to significantly save the computation in magnetic resonance spectroscopy and imaging [25, 53, 54]. To further improve efficiency, lightweight deep learning structures can be obtained by network pruning [55], knowledge distillation [56], and neural architecture search [57].

## VI. CONCLUSION

In this work, for exponential function and biomedical signal reconstruction, we proposed a new approach, alternating deep low-rank (ADLR), which utilizes the optimization solver and deep learning alternately to solve the low-rank model. The analysis of intermediate variables of reconstructing magnetic resonance signals synthesized by exponentials indicates the successful utilization of the low-rank property.

Experimental results on biomedical signals show that the trained ADLR can faithfully reconstruct nuclear magnetic resonance spectra when the sampling rates mismatch between training and test dataset, or reconstruct magnetic resonance imaging signals of different organs. These performances indicate that the proposed ADLR can effectively alleviate the drop in reconstruction quality due to the mismatches of attributes of the dataset.

## ACKNOWLEDGMENT

The authors thank Vladislav Orekhov and Maxim Mayzel for sharing the NMR data. The authors also thank NMRPipe and SPARKY for data processing support.

## REFERENCES

[1] Y. Yu, A. P. Petropulu and H. V. Poor, "Mimo radar using compressive sampling," *IEEE Journal of Selected Topics in Signal Processing*, vol. 4, no. 1, pp. 146-163, 2010.
[2] A. Hirose and S. Yoshida, "Generalization characteristics of complex-valued feedforward neural networks in relation to signal coherence," *IEEE Trans. Neural Netw. Learn Syst.*, vol. 23, no. 4, pp. 541-551, 2012.
[3] J. A. Tropp, J. N. Laska, M. F. Duarte, J. K. Romberg and R. G. Baraniuk, "Beyond nyquist: Efficient sampling of sparse bandlimited signals," *IEEE Transactions on Information Theory*, vol. 56, no. 1, pp. 520-544, 2010.
[4] M. Hansson, T. Gansler and G. Salomonsson, "Estimation of single event-related potentials utilizing the Prony method," *IEEE Trans. Biomed. Eng.*, vol. 43, no. 10, pp. 973-981, 1996.
[5] Q. Yang, Z. Wang, K. Guo, C. Cai and X. Qu, "Physics-driven synthetic data learning for biomedical magnetic resonance: The imaging physics-based data synthesis paradigm for artificial intelligence," *IEEE Signal Process. Mag.*, vol. 40, no. 2, pp. 129-140, 2023.
[6] F. Bloch, "Nuclear induction," *Phys. Rev.*, vol. 70, no. 7-8, p. 460, 1946.
[7] J. C. Hoch and A. S. Stern, *NMR Data Processing*. Wiley, 1996.
[8] D. Guo, H. Lu and X. Qu, "A fast low rank Hankel matrix factorization reconstruction method for non-uniformly sampled magnetic resonance spectroscopy," *IEEE Access*, vol. 5, pp. 16033-16039, 2017.
[9] X. Qu, M. Mayzel, J.-F. Cai, Z. Chen and V. Orekhov, "Accelerated NMR spectroscopy with low-rank reconstruction," *Angew. Chem.-Int. Edit.*, vol. 54, no. 3, pp. 852-854, 2015.
[10] J. Ying, H. Lu, Q. Wei, J.-F. Cai, D. Guo, J. Wu, Z. Chen and X. Qu, "Hankel matrix nuclear norm regularized tensor completion for N-dimensional exponential signals," *IEEE Trans. Signal Process.*, vol. 65, no. 14, pp. 3702-3717, 2017.
[11] P. Koehl, "Linear prediction spectral analysis of NMR data," *Prog. Nucl. Magn. Reson. Spectrosc.*, vol. 34, no. 3-4, pp. 257-299, 1999.
[12] Z. Wang, D. Guo, Z. Tu, Y. Huang, Y. Zhou, J. Wang, L. Feng, D. Lin, Y. You, T. Agback, V. Orekhov and X. Qu, "A sparse model-inspired deep thresholding network for exponential signal reconstruction-application in fast biological spectroscopy," *IEEE Trans. Neural Netw. Learn. Syst.*, vol. 34, no. 10, pp. 7578 - 7592, 2023.
[13] Z. Wang, C. Qian, D. Guo, H. Sun, R. Li, B. Zhao and X. Qu, "One-dimensional deep low-rank and sparse network for accelerated MRI," *IEEE Trans. Med. Imaging*, vol. 42, no. 1, pp. 79-90, 2022.

[14] X. Zhang, H. Lu, D. Guo, Z. Lai, H. Ye, X. Peng, B. Zhao and X. Qu, "Accelerated MRI reconstruction with separable and enhanced low-rank Hankel regularization," *IEEE Trans. Med. Imaging,* vol. 41, no. 9, pp. 2486-2498, 2022.
[15] J. P. Haldar, "Low-rank modeling of local k-space neighborhoods (LORAKS) for constrained MRI," *IEEE Trans. Med. Imaging,* vol. 33, no. 3, pp. 668-681, 2014.
[16] H. K. Aggarwal, M. P. Mani and M. Jacob, "MoDL: Model-based deep learning architecture for inverse problems," *IEEE Trans. Med. Imaging,* vol. 38, no. 2, pp. 394-405, 2019.
[17] H. M. Nguyen, X. Peng, M. N. Do and Z.-P. Liang, "Denoising mr spectroscopic imaging data with low-rank approximations," *IEEE Trans. Biomed. Eng.,* vol. 60, no. 1, pp. 78-89, 2013.
[18] S. J. Nelson, "Magnetic resonance spectroscopic imaging," *IEEE Eng. Med. Biol. Mag.,* vol. 23, no. 5, pp. 30-39, 2004.
[19] X. Qu, Y. Huang, H. Lu, T. Qiu, D. Guo, T. Agback, V. Orekhov and Z. Chen, "Accelerated nuclear magnetic resonance spectroscopy with deep learning," *Angew. Chem.-Int. Edit.,* vol. 59, no. 26, pp. 10297-10300, 2020.
[20] Y. Huang, J. Zhao, Z. Wang, V. Orekhov, D. Guo and X. Qu, "Exponential signal reconstruction with deep Hankel matrix factorization," *IEEE Trans. Neural Netw. Learn Syst.,* vol. 34, no. 9, pp. 6214 - 6226, 2023.
[21] X. Qu, X. Cao, D. Guo and Z. Chen, "Compressed sensing for sparse magnetic resonance spectroscopy," in *Proc. Int. Soc. Magn. Reson. Med. Sci. Meeting (ISMRM)*, 2010, p. 3371.
[22] K. Kazimierczuk and V. Y. Orekhov, "Accelerated NMR spectroscopy by using compressed sensing," *Angew. Chem.-Int. Edit.,* vol. 50, no. 24, pp. 5556-5559, 2011.
[23] T. Qiu, Z. Wang, H. Liu, D. Guo and X. Qu, "Review and prospect: NMR spectroscopy denoising and reconstruction with low-rank Hankel matrices and tensors," *Magn. Reson. Chem.,* vol. 59, no. 3, pp. 324-345, 2021.
[24] K. Yao, E. Lin, M. Liu, X. Qu, Y. Yang, K. Song, J. Xie, H. Sun and Z. Chen, "Accelerated detection for low-field NMR using non-uniform sampling and improved reconstruction," *IEEE Transactions on Instrumentation and Measurement,* vol. 71, pp. 1-11, 2022.
[25] J. Ying, J.-F. Cai, D. Guo, G. Tang, Z. Chen and X. Qu, "Vandermonde factorization of Hankel matrix for complex exponential signal recovery—application in fast NMR spectroscopy," *IEEE Trans. Signal Process.,* vol. 66, no. 21, pp. 5520-5533, 2018.
[26] J.-F. Cai, E. J. Candes and Z. Shen, "A singular value thresholding algorithm for matrix completion," *SIAM J. Optim.,* vol. 20, no. 4, pp. 1956-1982, 2010.
[27] N. Srebro, "Learning with matrix factorizations," Ph.D. dissertation, Massachusetts Institute of Technology, 2004.
[28] Y. LeCun, Y. Bengio and G. Hinton, "Deep learning," *Nature,* vol. 521, no. 7553, pp. 436-444, 2015.
[29] E. Tjoa and C. Guan, "A survey on explainable artificial intelligence (xai): Toward medical xai," *IEEE Trans. Neural Netw. Learn Syst.,* vol. 32, no. 11, pp. 4793-4813, 2021.
[30] Z. Li, F. Liu, W. Yang, S. Peng and J. Zhou, "A survey of convolutional neural networks: Analysis, applications, and prospects," *IEEE Trans. Neural Netw. Learn. Syst.,* vol. 33, no. 12, pp. 6999-7019, 2022.
[31] G. Karunanithy and D. F. Hansen, "FID-Net: A versatile deep neural network architecture for NMR spectral reconstruction and virtual decoupling," *J. Biomol. NMR,* vol. 75, no. 4, pp. 179-191, 2021.
[32] V. Antun, F. Renna, C. Poon, B. Adcock and A. C. Hansen, "On instabilities of deep learning in image reconstruction and the potential costs of AI," *Proc. Natl. Acad. Sci. U S A,* vol. 117, no. 48, pp. 30088-30095, 2020.
[33] M. Blaimer, M. Gutberlet, P. Kellman, F. A. Breuer, H. Köstler and M. A. Griswold, "Virtual coil concept for improved parallel MRI employing conjugate symmetric signals," *Magn. Reson. Med.,* vol. 61, no. 1, pp. 93-102, 2009.
[34] S. Boyd, N. Parikh, E. Chu, B. Peleato and J. Eckstein, "Distributed optimization and statistical learning via the alternating direction method of multipliers," *Found. Trends Mach. Learn.,* vol. 3, no. 1, pp. 1-122, 2011.
[35] G. Huang, Z. Liu, L. Van Der Maaten and K. Q. Weinberger, "Densely connected convolutional networks," in *Proc. IEEE Conf. Comput. Vis. Pattern Recognit. (CVPR)*, 2017, pp. 4700-4708.
[36] Ö. Yeniay, "Penalty function methods for constrained optimization with genetic algorithms," *Math Comput. Appl.,* vol. 10, no. 1, pp. 45-56, 2005.
[37] V. Nair and G. E. Hinton, "Rectified linear units improve restricted boltzmann machines," in *Proc. Int. Conf. Mach. Learn. (ICML)*, 2010, pp. 807-814.
[38] S. Ioffe and C. Szegedy, "Batch normalization: Accelerating deep network training by reducing internal covariate shift," in *Proc. Int. Conf. Mach. Learn. (ICML)*, 2015, pp. 448-456.
[39] K. He, X. Zhang, S. Ren and J. Sun, "Deep residual learning for image recognition," in *Proc. IEEE Conf. Comput. Vis. Pattern Recognit. (CVPR)*, 2016, pp. 770-778.
[40] D. P. Kingma and J. Ba, "Adam: A method for stochastic optimization," *arXiv:1412.6980.* 2014.
[41] P. J. Shin, P. E. Z. Larson, M. A. Ohliger, M. Elad, J. M. Pauly, D. B. Vigneron and M. Lustig, "Calibrationless parallel imaging reconstruction based on structured low-rank matrix completion," *Magn. Reson. Med.,* vol. 72, no. 4, pp. 959-970, 2014.
[42] J. Wright, A. Ganesh, S. Rao, Y. Peng and Y. Ma, "Robust principal component analysis: Exact recovery of corrupted low-rank matrices via convex optimization," in *Proc. Adv. Neural Inf. Process. Syst.(NIPS)*, 2009, vol. 22, pp. 2080-2088.
[43] M. Gal, M. Mishkovsky and L. Frydman, "Real-time monitoring of chemical transformations by ultrafast 2D NMR spectroscopy," *J. Am. Chem. Soc.,* vol. 128, no. 3, pp. 951-956, 2006.
[44] T. W. M. Fan and A. N. Lane, "Applications of NMR spectroscopy to systems biochemistry," *Prog. Nucl. Magn. Reson. Spectrosc.,* vol. 92-93, pp. 18-53, 2016.
[45] S. G. Hyberts, K. Takeuchi and G. Wagner, "Poisson-gap sampling and forward maximum entropy reconstruction for enhancing the resolution and sensitivity of protein NMR data," *J. Am. Chem. Soc.,* vol. 132, no. 7, pp. 2145-2147, 2010.
[46] M. G. Harisinghani, A. O'Shea and R. Weissleder, "Advances in clinical MRI technology," *Sci. Transl. Med.,* vol. 11, no. 523, p. 2591, 2019.
[47] K. Hammernik, T. Klatzer, E. Kobler, M. P. Recht, D. K. Sodickson, T. Pock and F. Knoll, "Learning a variational network for reconstruction of accelerated MRI data," *Magn. Reson. Med.,* vol. 79, no. 6, pp. 3055-3071, 2018.
[48] J. Zbontar, F. Knoll, A. Sriram, T. Murrell, Z. Huang, M. J. Muckley, A. Defazio, R. Stern, P. Johnson and M. Bruno, "fastMRI: An open dataset and benchmarks for accelerated MRI," *arXiv:1811.08839.* 2018.
[49] T. Zhang, J. M. Pauly, S. S. Vasanawala and M. Lustig, "Coil compression for accelerated imaging with cartesian sampling," *Magn. Reson. Med.,* vol. 69, no. 2, pp. 571-582, 2013.
[50] J. P. Haldar and J. Zhuo, "P-LORAKS: Low-rank modeling of local k-space neighborhoods with parallel imaging data," *Magn. Reson. Med.,* vol. 75, no. 4, pp. 1499-1514, 2016.
[51] A. Pramanik, H. Aggarwal and M. Jacob, "Deep generalization of structured low-rank algorithms (Deep-SLR)," *IEEE Trans. Med. Imaging,* vol. 39, no. 12, pp. 4186-4197, 2020.
[52] J. Wu, R. Xu, Y. Huang, J. Zhan, Z. Tu, X. Qu and D. Guo, "Fast NMR spectroscopy reconstruction with a sliding window based Hankel matrix," *J. Magn. Reson.,* vol. 342, p. 107283, 2022.
[53] B. Zhao, K. Setsompop, E. Adalsteinsson, B. Gagoski, H. Ye, D. Ma, Y. Jiang, P. Ellen Grant, M. A. Griswold and L. L. Wald, "Improved magnetic resonance fingerprinting reconstruction with low‐rank and subspace modeling," *Magn. Reson. Med.,* vol. 79, no. 2, pp. 933-942, 2018.
[54] C. Ma, B. Clifford, Y. Liu, Y. Gu, F. Lam, X. Yu and Z. P. Liang, "High‐resolution dynamic 31P‐MRSI using a low‐rank tensor model," *Magn. Reson. Med.,* vol. 78, no. 2, pp. 419-428, 2017.
[55] S. Han, H. Mao and W. J. Dally, "Deep compression: Compressing deep neural networks with pruning, trained quantization and Huffman coding," in *Proc. Int. Conf. Learn. Represent. (ICLR)*, 2016, pp. 1-14.
[56] G. Hinton, O. Vinyals and J. Dean, "Distilling the knowledge in a neural network," *arXiv:1503.02531.* 2015.
[57] B. Zoph and Q. V. Le, "Neural architecture search with reinforcement learning," in *Proc. Int. Conf. Learn. Represent. (ICLR)*, 2017, pp. 1-16.

# Supplementary Materials For
# "Alternating Deep Low-Rank Approach for Exponential Function Reconstruction and Its Biomedical Magnetic Resonance Applications"

Yihui Huang, Zi Wang, Xinlin Zhang, Jian Cao, Zhangren Tu, Meijin Lin, Di Guo, Xiaobo Qu*

*S1. Parameters of Synthetic Exponential Functions*

TABLE S1
SYNTHETIC SIGNAL USED IN FIG. 1

| Peaks ID<br>Parameters | 1 | 2 | 3 | 4 | 5 |
|---|---|---|---|---|---|
| Amplitude($A$) | 0.300 | 0.475 | 0.650 | 0.825 | 1.000 |
| Damping factor($\tau$) | 50 | 75 | 100 | 125 | 150 |
| Phase($\phi$) | 0.4π | 0.8π | 1.2π | 1.6π | 2.0π |
| Frequency($f$) | 0.165 | 0.333 | 0.498 | 0.667 | 0.831 |
| Standard deviation of noise ($\sigma$) | 0.03 | | | | |

TABLE S2
SYNTHETIC SIGNAL USED IN FIG. 4 TO FIG. 7 AND FIG. 13 TO FIG. 17

| Peaks ID<br>Parameters | 1 | 2 | 3 | 4 | 5 |
|---|---|---|---|---|---|
| Amplitude($A$) | 0.100 | 0.325 | 0.550 | 0.775 | 1.000 |
| Damping factor($\tau$) | 50 | 75 | 100 | 125 | 150 |
| Phase($\phi$) | 0.4π | 0.8π | 1.2π | 1.6π | 2.0π |
| Frequency($f$) | 0.165 | 0.333 | 0.498 | 0.667 | 0.831 |
| Standard deviation of noise ($\sigma$) | 0.03 | | | | |

*S2. Experiment Setups of NMR Spectra*

*S2.1 Experiments Setup*

Four 2D NMR spectra were used in our reconstruction experiments. The detailed experiments setup of spectra are described as follows:

The 2D $^1$H–$^{15}$N HSQC spectrum (Fig. 6(a)) of GB1 was the data courtesy of Drs. Luke Arbogast and Frank Delaglio at National Institute of Standards and Technology, Institute for Bioscience and Biotechnology Research, USA. The sample was 2 mM U-$^{15}$N, 20%-$^{13}$C GB1 in 25 mM PO4, pH 7.0 with 150 mM NaCl and 5% D$_2$O. Data was collected using a phase-cycle selected HSQC at 298 K on a Bruker Advance 600 MHz spectrometer using a room temp HCN TXI probe, equipped with a z-axis gradient system. The fully sampled spectrum to be reconstructed consists of 1466×169 complex points, the direct dimension ($^1$H) has 1466 data points while the indirect dimension ($^{15}$N) has 169 data points.

The 2D $^1$H–$^{15}$N HSQC spectrum (Fig. S1(a)) of cytosolic CD79b protein was described in our previous work[S1]. In brief, the spectrum was acquired for 300 μM $^{15}$N-$^{13}$C labeled sample of cytosolic CD79b in 20 mM sodium phosphate buffer, pH 6.7 at 25 °C on 800 MHz Bruker AVANCE III HD spectrometer equipped with 3 mm TCI cryoprobe. The fully sampled spectrum to be reconstructed consists of 116×255 complex points, the direct dimension ($^1$H) has 116 data points while the indirect dimension ($^{15}$N) has 255 data points.

The 2D $^1$H–$^{15}$N HSQC spectrum (Fig. S2(a)) of Ubiquitin was acquired from ubiquitin sample at 298.2K temperature on an 800 MHz Bruker spectrometer and was described in the previous paper[S2]. The fully sampled spectrum to be reconstructed consists of 576×97 complex points, the direct dimension ($^1$H) has 576 data points while the indirect dimension ($^{15}$N) has 97 data points.

The 2D $^1$H–$^{15}$N best-TROSY spectrum (Fig. S3(a)) of ubiquitin was acquired at 298.2K temperature on an 800 MHz Bruker spectrometer and was described in the previous paper[S2]. The fully sampled spectrum to be reconstructed consists of 512×127 complex points, the direct dimension ($^1$H) has 512 data points while the indirect dimension ($^{15}$N) has 127 data points. undersampling retrospective with sampling rates of 25% and 50%.

*S2.2 Reconstructed Spectra*

To demonstrate the robustness of the trained ADLR model with sampling rates mismatch, the other three 2D spectra (Fig .S2-S4) are undersampling retrospectively with sampling rates of 25% and 50%. The state-of-the-art method CS[S3], LRHMF[S4] based on optimization, and FID-Net[S5], DHMF[S6] based on deep learning are compared. Consistent with the result in the main

> REPLACE THIS LINE WITH YOUR PAPER IDENTIFICATION NUMBER (DOUBLE-CLICK HERE TO EDIT) <     2text, trained DHMF cannot reconstruct the spectra with higher sampling rates stably, such as bringing more artifacts (Fig. S3(f)) and even failing to reconstruct (Fig. S1(f)). In some cases, CS obtains a poor reconstruction of the NMR spectra (Fig. S1(b)). FID-Net loses the low-intensity peaks (Fig. S1(d)) and generates some artifacts (Fig. S2(d)). On the contrary, the LRHMF and proposed ADLR can still reconstruct NMR spectra with high Pearson correlation ($r^2$>0.99) at a sampling rate of 50%.

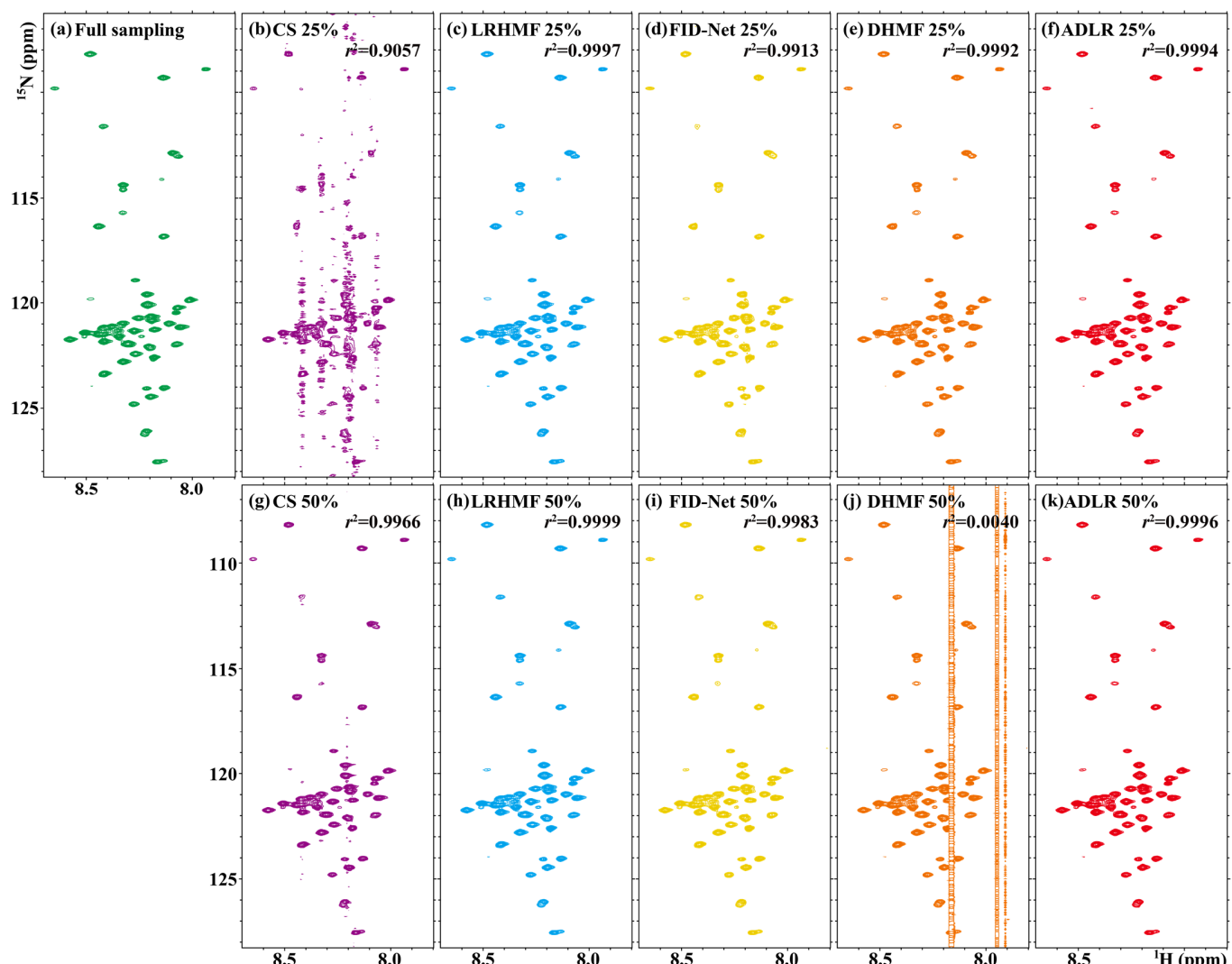

Fig. S1. The reconstructions of 2D $^1$H-$^{15}$N HSQC spectrum of cytosolic CD79b protein under the sampling rate mismatch. (a) The fully sampling. The CS, LRHMF, FID-Net, DHMF and ADLR trained with sampling rate 25% reconstruct the spectrum from (b-f) 25% data and (g-k) 50% data, respectively.

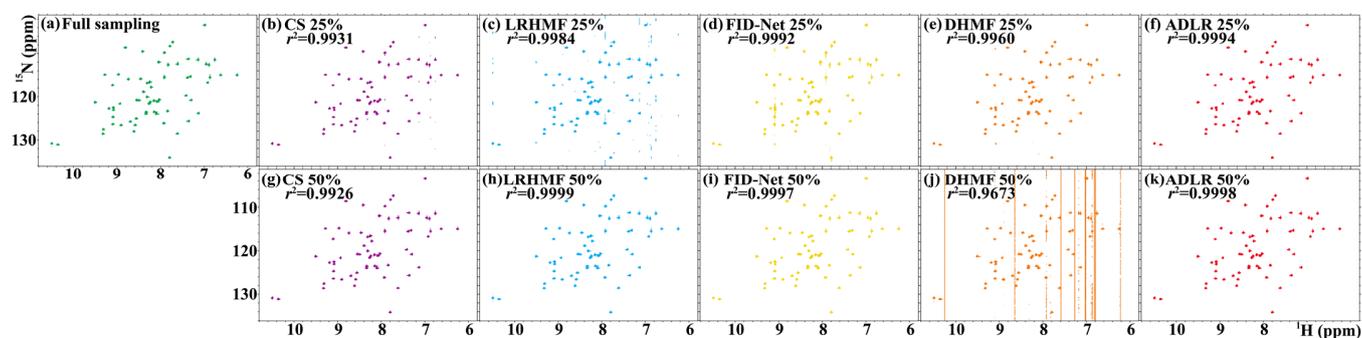

Fig. S2. The reconstructions of 2D $^1$H-$^{15}$N HSQC spectrum of Gb1 under the sampling rate mismatch. (a) The fully sampling. The CS, LRHMF, FID-Net, DHMF and ADLR trained with sampling rate 25% reconstruct the spectrum from (b-f) 25% data and (g-k) 50% data, respectively.

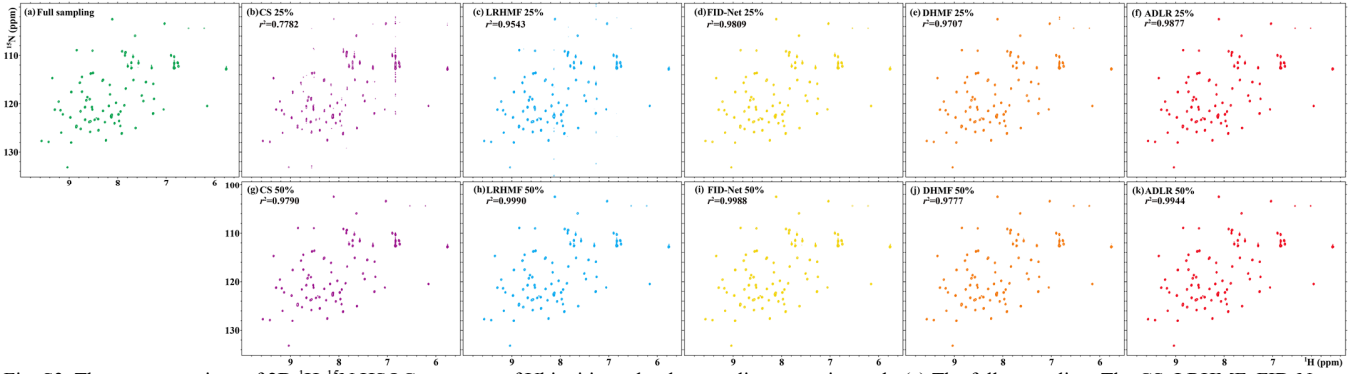

Fig. S3. The reconstructions of 2D $^1$H-$^{15}$N HSQC spectrum of Ubiquitin under the sampling rate mismatch. (a) The fully sampling. The CS, LRHMF, FID-Net, DHMF and ADLR trained with sampling rate 25% reconstruct the spectrum from (b-f) 25% data and (g-k) 50% data, respectively.

*S3. Effect of Different Parameters*

In the proposed ADLR, all regularization parameters are learnable, whose initial values and learned values are listed in Tables S3 and S4 for exponential functions and MRI reconstruction. The main non-learnable parameters include the number $K$ of iteration blocks, the number $L$ of layers in each deep learning solver and the weight $\alpha$ in the loss function. Under $K$ from 8 to 12, $L$ from 3 to 7 and $\alpha$ from 0 to $10^{-2}$, all RLNEs achieve about 0.025 from 25% data. Thus, these three parameters are set as 10, 6 and $10^{-2}$ respectively (the black arrows in Fig. S4). Due to the larger size 224×4 of multi-coil MRI k-space signals than the length 255 of exponential functions in training, the number of iteration blocks is reduced to 5 for MRI reconstruction and the other non-learnable parameters are the same.

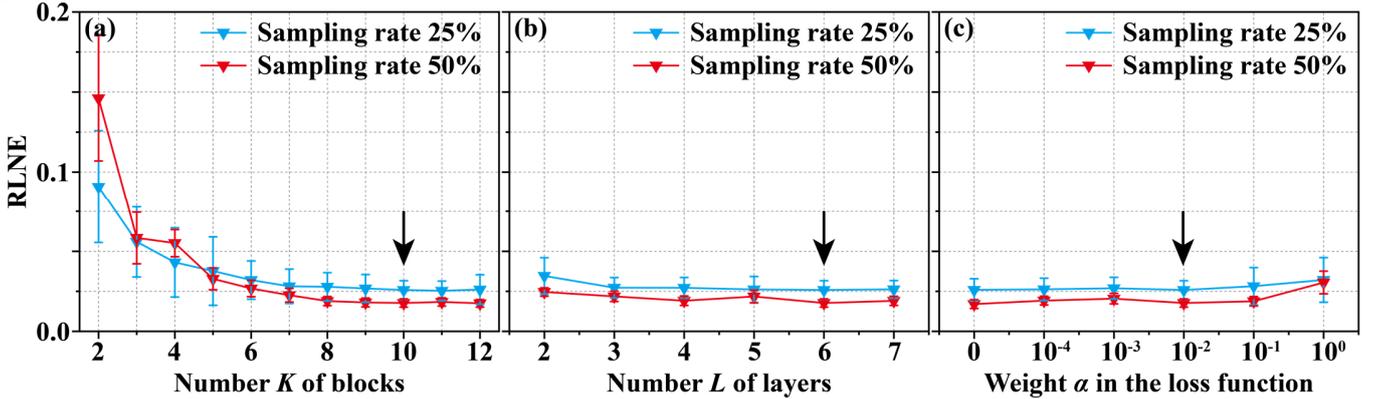

Fig. S4. Reconstruction error of ADLR under sampling rates of 25% and 50% with different (a) numbers $K$ of blocks, (c) numbers $L$ of layers and (c) weights $\alpha$ in the loss function.

For the reconstruction of synthetic exponential functions and NMR spectroscopy, FID-Net [S5] is the end-to-end model without regularization parameters, and the solved model of CS [S3] and LRHMF [S4] are defined:

$$\text{CS: } \min_{\mathbf{x}} \frac{1}{2}\|\mathbf{y}-\mathcal{U}\mathbf{x}\|_2^2 + \lambda\|\mathcal{F}\mathbf{x}\|_1, \tag{S1}$$

$$\text{LRHMF: } \min_{\mathbf{x},\mathbf{P},\mathbf{Q},\mathbf{D}} \frac{1}{2}(\|\mathbf{P}\|_F^2+\|\mathbf{Q}\|_F^2)+\frac{\lambda}{2}\|\mathbf{y}-\mathcal{U}\mathbf{x}\|_2^2+\frac{\beta}{2}\|\mathcal{H}\mathbf{x}-\mathbf{PQ}^H\|_F^2+\langle\mathbf{D},\mathcal{H}\mathbf{x}-\mathbf{PQ}^H\rangle. \tag{S2}$$

Table S3 Regularization parameters of methods for the reconstruction of synthetic exponential functions and NMR spectroscopy

| | Sampling rates | | 10% | 15% | 20% | 25% | 30% | 35% | 40% | 45% | 50% |
|---|---|---|---|---|---|---|---|---|---|---|---|
| CS | $\lambda$ | | 0.17 | 0.10 | 0.08 | 0.05 | 0.05 | 0.05 | 0.04 | 0.04 | 0.03 |
| LRHMF | $\lambda/\beta$ | | $10^{6.0}/1$ | $10^{3.0}/1$ | $10^{2.5}/1$ | $10^{2.5}/1$ | $10^{2.5}/1$ | $10^{2.5}/1$ | $10^{2.5}/1$ | $10^{2.5}/1$ | $10^{2.0}/1$ |
| | $k^{\text{th}}$ block | | 1 | 2 | 3 | 4 | 5 | 6 | 7 | 8 | 9 | 10 |
| ADLR | Initial values | $\gamma_{DL}^k$ | $10^4$ | | | | | | | | | |
| | | $\gamma^k$ | $10^4$ | | | | | | | | | |
| | | $\beta_P^k$ | 100.0 | | | | | | | | | |
| | | $\beta_Q^k$ | 100.0 | | | | | | | | | |
| | Learned values | $\gamma_{DL}^k$ | $5.2\times10^5$ | $6.7\times10^4$ | $6.7\times10^4$ | $3.7\times10^1$ | $4.4\times10^0$ | $4.0\times10^0$ | $2.3\times10^0$ | $2.0\times10^0$ | $1.2\times10^0$ | $4.6\times10^{-1}$ |
| | | $\gamma^k$ | $1.9\times10^5$ | $1.3\times10^5$ | $1.6\times10^0$ | $5.4\times10^{-1}$ | $2.6\times10^{-1}$ | $1.4\times10^{-1}$ | $1.1\times10^{-1}$ | $9.2\times10^{-2}$ | $8.9\times10^{-2}$ | $9.5\times10^{-2}$ |
| | | $\beta_P^k$ | 96.0 | 98.4 | 98.5 | 99.5 | 99.0 | 99.2 | 98.5 | 99.4 | 100.0 | 98.4 |
| | | $\beta_Q^k$ | 87.6 | 97.0 | 98.9 | 95.7 | 102.1 | 102.7 | 105.7 | 105.0 | 107.6 | 109.3 |

For the reconstruction of multi-coil MRI images, we implement the P-LORAKS algorithm [S7] using the publicly available Matlab code package: https://mr.usc.edu/download/LORAKS2/. Besides, the regularization parameters of the deep-learning-based ODLS [S8] and HDSLR [S9] are set according to the corresponding paper.

Table S4 Regularization parameters of methods for the reconstruction of multi-coil MRI images

| Dataset | | Coronal PDw knee | Sagittal PDw knee | Axial T2w brain | | |
|---|---|---|---|---|---|---|
| **P-LORAKS** | Lambda | $10^{-5.0}$ | $10^{-5.0}$ | $10^{-5.0}$ | | |
| | Rank | 50 | 110 | 80 | | |
| | R | 3 | 3 | 4 | | |
| **ADLR** | $k^{th}$ block | 1 | 2 | 3 | 4 | 5 |
| | Initial values $\gamma_{DL}^k$ | $10^4$ | | | | |
| | $\gamma^k$ | $10^4$ | | | | |
| | $\beta_P^k$ | 100.0 | | | | |
| | $\beta_Q^k$ | 100.0 | | | | |
| | Learned values $\gamma_{DL}^k$ | $1.4\times10^5$ | $4.1\times10^5$ | $3.2\times10^5$ | $7.7\times10^3$ | $6.4\times10^4$ |
| | $\gamma^k$ | $2.1\times10^5$ | $1.1\times10^5$ | $2.4\times10^5$ | $2.0\times10^5$ | $3.4\times10^5$ |
| | $\beta_P^k$ | 92.3 | 100.5 | 101.9 | 103.0 | 98.1 |
| | $\beta_Q^k$ | 75.7 | 75.4 | 77.7 | 79.2 | 86.0 |

*S4. Measure the Mismatch between the Target and the Training Signals*

The 1-Wasserstein distance[S10], defined as the minimum cost of transport from one distribution to another, is adopted to measure the mismatch between the target and training signals. Details are provided below:

The target and training signals are represented by probability mass functions $p(\mathbf{x})$ and $q(\mathbf{x})$ of all corresponding undersampling signals $\mathbf{x}$. Except for zeros at undersampled positions, the other values are uniformly spaced in 100 intervals according to the range of magnitude values of synthetic exponentials and *in vivo* k-space signals. The logarithmic function is used firstly due to the large range of values in k-space signals.

The 1-Wasserstein distance of the probability mass functions $p(\mathbf{x})$ and $q(\mathbf{x})$ is defined as:

$$W(p(\mathbf{x}), q(\mathbf{x})) = \underset{\mathbf{T}}{\text{minimize}} <\mathbf{T}, \mathbf{C}>$$
$$\text{subject to } \sum_{i}^{101} \mathbf{T}_{ij} = p(\mathbf{x})_j \quad \forall j = \{1, \dots, 101\},$$
$$\sum_{j}^{101} \mathbf{T}_{ij} = q(\mathbf{x})_i \quad \forall i = \{1, \dots, 101\},$$
$$\mathbf{T}_{ij} \geq 0 \quad \forall i = \{1, \dots, 101\} \text{ and } \forall j = \{1, \dots, 101\}, \quad \text{(S3)}$$

where C is the cost matrix defined as $\mathbf{C}_{ij}=0$ if $i = j$, otherwise $\mathbf{C}_{ij}=1$, and $\mathbf{T}$ is the joint probability mass function with the marginal distributions $p(\mathbf{x})$ and $q(\mathbf{x})$.

For the training dataset of synthetic exponentials at a sampling rate of 25% (Table S5), the target signals with the matched sampling rate achieve the lowest Wasserstein distance among sampling rates from 10% to 50%. For the training dataset of coronal PDw knee image (Table S6), the target signals of the axial T2w brain achieve the highest distance due to the mismatch of the organ and imaging contrast.

TABLE S5
The 1-Wasserstein distance between the target signals with different sampling rates and training signals of the sampling rate of 25% in synthetic exponentials

| Sampling rates of target signals | 10% | 15% | 20% | 25% | 30% | 35% | 40% | 45% | 50% |
|---|---|---|---|---|---|---|---|---|---|
| Distance to training | 0.158 | 0.120 | 0.079 | **0.058** | 0.099 | 0.135 | 0.185 | 0.231 | 0.280 |

TABLE S6
The 1-Wasserstein distance between the target signals of different MRI image datasets and training signals of coronal PDw knee

| MRI images Dataset of target signals | Coronal PDw knee | Sagittal PDw knee | Axial T2w brain |
|---|---|---|---|
| Distance to training | **0.016** | 0.018 | 0.046 |

*S5. Reconstruction of MRI Images*

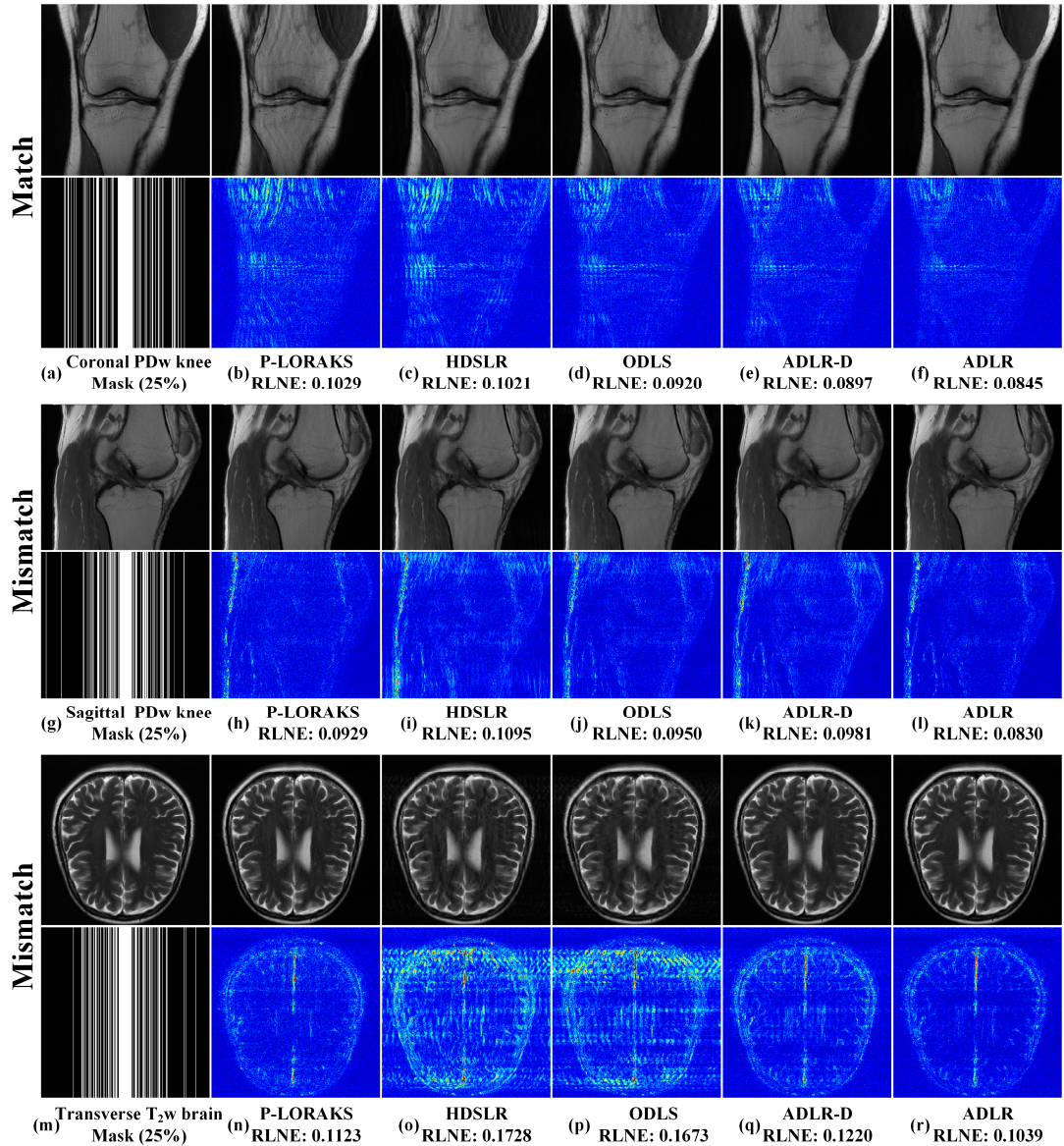

Fig. S5. Reconstructed images of three MRI datasets. The fully sampled images of (a) coronal PDw knee, (g) sagittal PDw knee, and (m) axial $T_2w$ brain and corresponding 1D Cartesian sampling patterns with a sampling rate of 25%. (b-f)(h-l)(n-r) The corresponding reconstructed images and error maps (5×) by P-LORAKS, HDSLR, ODLS, ADLR-D, and ADLR.

*S5. Comparison with DHMF*

The optimization solvers effectively constrain variables **P** and **Q** obtained by deep learning solvers. As the visualization of the variables **P** and **Q** illustrate, the ADLR provides the variables similar visually to the full sampling (Fig. S6(a-d)). Noticeably, variable **Q** contains some noise-like column vectors (Fig. S6(b)) but does not significantly affect the reconstruction result due to the small magnitudes of corresponding vectors of **P** (Fig. S6(a)). Besides, the subsequent low-rank optimization solver is independent of the data. Thus, when reconstructing the mismatched signals, the reconstruction performance of the deep learning solver drops but the optimization solver still works effectively, leading to the high generalization ability.

As a comparison, the reconstructed signal from DHMF is close to the full sampling (Fig. S6(h)), apparently similar to ADLR. However, without the constraint of optimization solvers, the variables **P** and **Q** in DHMF contain plenty of noise-like column vectors (Fig. S6(e-g)). In the absence of the correction of the following optimization solvers, DHMF has a low generalization ability for mismatched sampling rates.

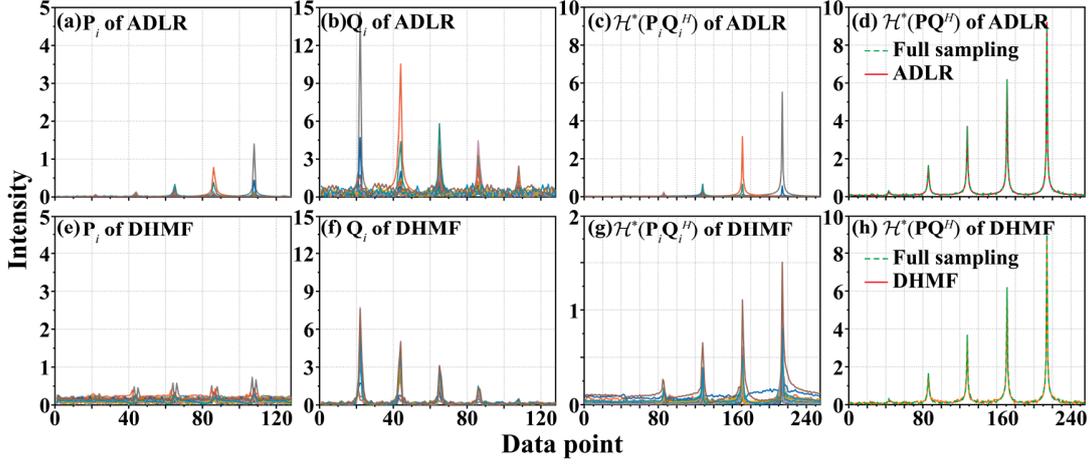

Fig. S6. The outputted variables of the 10$^{th}$ deep learning solver in the frequency domain. The 20 column vectors of variable (a) **P**, (b) **Q**, and (c) the product of corresponding column vectors after inverse Hankel operation in ADLR. (d) The summation of vectors of (c). (e-h) The corresponding vectors in DHMF.

*S5. Introduction of Optimization Solver*

For the constrained problem of reconstructing the signal **x** from the undersampled signal **y** under the undersampling operator $\mathcal{U}$:

$$\min_{\mathbf{x},\mathbf{P},\mathbf{Q}} \frac{1}{2}(\|\mathbf{P}\|_F^2+\|\mathbf{Q}\|_F^2)+\frac{\lambda}{2}\|\mathbf{y}-\mathcal{U}\mathbf{x}\|_2^2 \ s.t. \ \mathbf{P}\mathbf{Q}^H=\mathcal{H}\mathbf{x}, \tag{S4}$$

we utilize the penalty function model [S11] and convert Eq. (S4) to the unconstrained problem:

$$\min_{\mathbf{x},\mathbf{P},\mathbf{Q}} \frac{1}{2}(\|\mathbf{P}\|_F^2+\|\mathbf{Q}\|_F^2)+\frac{\lambda}{2}\|\mathbf{y}-\mathcal{U}\mathbf{x}\|_2^2+\frac{\beta}{2}\|\mathcal{H}\mathbf{x}-\mathbf{P}\mathbf{Q}^H\|_F^2, \tag{S5}$$

where $\beta>0$ is the penalty coefficient and $\lambda$ is the trade-off parameter between sampled signals and reconstructed signals. The minimization of Eq. (S5) can be solved by successively minimizing three sub-problems in the iterative scheme as follows:

$$\mathbf{P}^{k+1}=\arg\min_{\mathbf{P}}\frac{1}{2}(\|\mathbf{P}\|_F^2+\|\mathbf{Q}^k\|_F^2)+\frac{\lambda}{2}\|\mathbf{y}-\mathcal{U}\mathbf{x}^k\|_2^2+\frac{\beta}{2}\|\mathcal{H}\mathbf{x}^k-\mathbf{P}(\mathbf{Q}^k)^H\|_F^2, \tag{S6}$$

$$\mathbf{Q}^{k+1}=\arg\min_{\mathbf{Q}}\frac{1}{2}(\|\mathbf{P}^{k+1}\|_F^2+\|\mathbf{Q}\|_F^2)+\frac{\lambda}{2}\|\mathbf{y}-\mathcal{U}\mathbf{x}^k\|_2^2+\frac{\beta}{2}\|\mathcal{H}\mathbf{x}^k-\mathbf{P}^{k+1}\mathbf{Q}^H\|_F^2, \tag{S7}$$

$$\mathbf{x}^{k+1}=\arg\min_{\mathbf{x}}\frac{1}{2}(\|\mathbf{P}^{k+1}\|_F^2+\|\mathbf{Q}^{k+1}\|_F^2)+\frac{\lambda}{2}\|\mathbf{y}-\mathcal{U}\mathbf{x}\|_2^2+\frac{\beta}{2}\|\mathcal{H}\mathbf{x}-\mathbf{P}^{k+1}(\mathbf{Q}^{k+1})^H\|_F^2. \tag{S8}$$

1) Update **P** and **Q**

To update the variable **P**, we fix the value of the variables $\mathbf{Q}^k$ and $\mathbf{x}^k$, and the sub-problem (S6) can be rewritten below:

$$\mathbf{P}^{k+1}=\arg\min_{\mathbf{P}}\frac{1}{2}\|\mathbf{P}\|_F^2+\frac{\beta}{2}\|\mathcal{H}\mathbf{x}^k-\mathbf{P}(\mathbf{Q}^k)^H\|_F^2. \tag{S9}$$

This is a least squares problem and the solution is:

$$\mathbf{P}^{k+1}=\beta\mathcal{H}\mathbf{x}^k\mathbf{Q}^k\left(\mathbf{I}+\beta(\mathbf{Q}^k)^H\mathbf{Q}^k\right)^{-1}. \tag{S10}$$

Similar to the update of the variable **P**, **Q** is solved with the fixed variables $\mathbf{P}^{k+1}$ and $\mathbf{x}^k$ according to the following equation:

$$\mathbf{Q}^{k+1}=\arg\min_{\mathbf{Q}}\frac{1}{2}\|\mathbf{Q}\|_F^2+\frac{\beta}{2}\|\mathcal{H}\mathbf{x}^k-\mathbf{P}^{k+1}\mathbf{Q}^H\|_F^2, \tag{S11}$$

$$\mathbf{Q}^{k+1}=\beta(\mathcal{H}\mathbf{x}^k)^H\mathbf{P}^{k+1}(\mathbf{I}+\beta(\mathbf{P}^{k+1})^H\mathbf{P}^{k+1})^{-1}. \tag{S12}$$

2) Update **x**

For the fixed $\mathbf{P}^{k+1}$ and $\mathbf{Q}^{k+1}$, the sub-problem (S8) can be rewritten as:

$$\mathbf{x}^{k+1}=\arg\min_{\mathbf{x}}\frac{\lambda}{2}\|\mathbf{y}-\mathcal{U}\mathbf{x}\|_2^2+\frac{\beta}{2}\|\mathcal{H}\mathbf{x}-\mathbf{P}^{k+1}(\mathbf{Q}^{k+1})^H\|_F^2. \tag{S13}$$

The solution is:

$$\mathbf{x}^{k+1}=(\beta\mathcal{H}^*\mathcal{H}+\lambda\mathcal{U}^*\mathcal{U})^{-1}(\lambda\mathcal{U}^*\mathbf{y}+\beta\mathcal{H}^*(\mathbf{P}^{k+1}(\mathbf{Q}^{k+1})^H)). \tag{S14}$$

Finally, the whole optimization solver can be written as:

$$\begin{cases} \mathbf{P}^{k+1}=\beta\mathcal{H}\mathbf{x}^k\mathbf{Q}^k\left(\mathbf{I}+\beta(\mathbf{Q}^k)^H\mathbf{Q}^k\right)^{-1} \\ \mathbf{Q}^{k+1}=\beta(\mathcal{H}\mathbf{x}^k)^H\mathbf{P}^{k+1}(\mathbf{I}+\beta(\mathbf{P}^{k+1})^H\mathbf{P}^{k+1})^{-1} \\ \mathbf{x}^{k+1}=(\beta\mathcal{H}^*\mathcal{H}+\lambda\mathcal{U}^*\mathcal{U})^{-1}(\lambda\mathcal{U}^*\mathbf{y}+\beta\mathcal{H}^*(\mathbf{P}^{k+1}(\mathbf{Q}^{k+1})^H)) \end{cases}. \tag{S15}$$

Here, the penalty function model is adopted instead of the widely-used augmented Lagrange model because the former has only three sub-problems to solve, while the latter has four sub-problems due to the added Lagrange multiplier. Besides, all regularization parameters, i.e. $\lambda$ and $\beta$, are learnable.

*S6. Effect of Changing the Learning Rate*

A large learning rate leads to large increments in the updated weights, which is a possible reason why the RLNE of the validation dataset does not decrease steadily (epochs 35 to 52 in Fig. S7(b)). In comparison, at the end of the training, the reduced learning rate can alleviate the oscillation of the RLNE of the validation dataset and achieve a lower RLNE steadily (Fig. S7(a)).

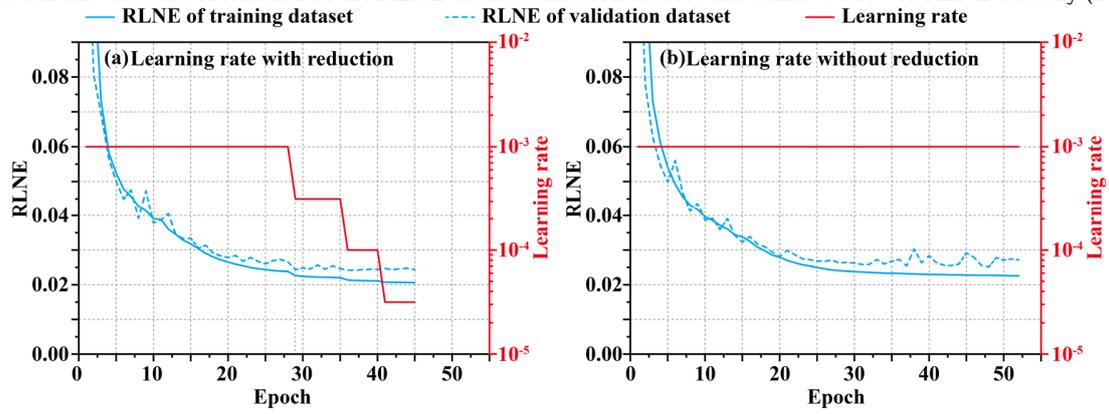

Fig. S7. The RLNEs of the training and validation dataset with (a) the reduced learning rate and (b) the constant learning rate over the epoch.

# References

[S1] X. Qu, M. Mayzel, J.-F. Cai, Z. Chen and V. Orekhov, "Accelerated NMR spectroscopy with low-rank reconstruction," *Angew. Chem.-Int. Edit.,* vol. 54, no. 3, pp. 852-854, 2015.

[S2] M. Mayzel, K. Kazimierczuk and V. Y. Orekhov, "The causality principle in the reconstruction of sparse NMR spectra," *Chemical Communications,* vol. 50, no. 64, pp. 8947-8950, 2014.

[S3] K. Kazimierczuk and V. Y. Orekhov, "Accelerated NMR spectroscopy by using compressed sensing," *Angew. Chem.-Int. Edit.,* vol. 50, no. 24, pp. 5556-5559, 2011.

[S4] D. Guo, H. Lu and X. Qu, "A fast low rank Hankel matrix factorization reconstruction method for non-uniformly sampled magnetic resonance spectroscopy," *IEEE Access,* vol. 5, pp. 16033-16039, 2017.

[S5] G. Karunanithy and D. F. Hansen, "FID-Net: A versatile deep neural network architecture for NMR spectral reconstruction and virtual decoupling," *J. Biomol. NMR,* vol. 75, no. 4, pp. 179-191, 2021.

[S6] Y. Huang, J. Zhao, Z. Wang, V. Orekhov, D. Guo and X. Qu, "Exponential signal reconstruction with deep Hankel matrix factorization," *IEEE Trans. Neural Netw. Learn Syst.,* 2021.

[S7] J. P. Haldar and J. Zhuo, "P-LORAKS: Low-rank modeling of local k-space neighborhoods with parallel imaging data," *Magn. Reson. Med.,* vol. 75, no. 4, pp. 1499-1514, 2016.

[S8] Z. Wang, C. Qian, D. Guo, H. Sun, R. Li, B. Zhao and X. Qu, "One-dimensional deep low-rank and sparse network for accelerated MRI," *IEEE Trans. Med. Imaging,* vol. 42, no. 1, pp. 79-90, 2022.

[S9] A. Pramanik, H. Aggarwal and M. Jacob, "Deep generalization of structured low-rank algorithms (Deep-SLR)," *IEEE Trans. Med. Imaging,* vol. 39, no. 12, pp. 4186-4197, 2020.

[S10] G. Peyré and M. Cuturi, "Computational optimal transport: With applications to data science," *Found. Trends Mach. Learn.,* vol. 11, no. 5-6, pp. 355-607, 2019.

[S11] Ö. Yeniay, "Penalty function methods for constrained optimization with genetic algorithms," *Math Comput. Appl.,* vol. 10, no. 1, pp. 45-56, 2005.